\UseRawInputEncoding
\documentclass[12pt]{iopart}

\usepackage{graphicx}
\usepackage{dcolumn}
\usepackage{bm}
\usepackage{hyperref}

\usepackage{graphicx}
\usepackage{dcolumn}
\usepackage{bm}
\usepackage{hyperref}

\usepackage[caption=false]{subfig}
\usepackage{iopams} 
\usepackage{cite} 
\usepackage{xcolor} 

\usepackage{etoolbox}
\makeatletter
\newcommand{\mainmatter}{%
	\setcounter{footnote}{0}%
	\patchcmd{\@makefntext}{\fnsymbol}{\arabic}{}{}%
	\patchcmd{\@thefnmark}{\fnsymbol}{\arabic}{}{}%
	\def\@makefnmark{\textsuperscript{\arabic{footnote}}}%
}
\makeatother
\mainmatter

\def\ra{\rangle}
\def\la{\langle}\def\rmd{\mathrm{d}}  

\newcommand{\be}{\begin{eqnarray}}
\newcommand{\ee}{\end{eqnarray}}
\newcommand{\beq}{\begin{equation}}
\newcommand{\eeq}{\end{equation}}
\newcommand{\kmps}{\,\mathrm{km}\,\mathrm{s}^{-1}} 
\newcommand{\exclude}[1]{}

\graphicspath{{./Figures/}} 

\begin{document}
       \title{The Pierre Auger  Exotic  Events   and    Axion Quark Nuggets }
       \author{  Ariel  Zhitnitsky}
       \address{Department of Physics and Astronomy, University of British Columbia, Vancouver, V6T 1Z1, BC, Canada}
     
      \begin{abstract}
          The Pierre Auger Observatory  have reported  \cite{PierreAuger:2021int,2019EPJWC.19703003C,Colalillo:2017uC}  observation of several  exotic cosmic ray  -like events which apparently related to thunderstorms. These events are much larger in size than conventional cosmic ray  events, and they have very distinct 
     timing features.  
     A  possible nature of the observed phenomenon is still a matter of active research and debates
     as many   unusual features of these exotic events   are hard to explain.  
     In particular,  the frequency of appearance of these  exotic events    is very low (less than 2 events/year), in huge contrast with a typical rate of a conventional   lightning strikes in the area.   We propose that the observed  exotic events 
     can be explained  within the so-called axion quark nugget (AQN) dark matter   model. 
     The idea is that the AQNs    may trigger and initiate  a special and unique class of   lightning strikes  during a  thunderstorm as a result of  ionization of the  atmospheric molecules along its path. The corresponding AQN-induced lighting flashes  may  show   some specific features not shared by typical and much more frequent conventional   flashes. 
       We  support this proposal by demonstrating that  the  observations \cite{PierreAuger:2021int,2019EPJWC.19703003C,Colalillo:2017uC}, including the    frequency of appearance  and  time duration are consistent with observations.   We also comment on possible relation of AUGER exotic events with the Telescope Array bursts and the  terrestrial gamma ray flashes.   We list a number of features   of the AQN-induced  exotic events (such as specific radio pulses synchronized with  these events) which can be directly tested by  future experiments. We also suggest to use distributed acoustic sensing   instruments to detect the acoustic pulses which must be  synchronized  with AUGER exotic events. 
        \end{abstract}

\submitto{\jpg}
	\maketitle
	
\section{Introduction}\label{sec:introduction}

   The  AUGER   collaboration have reported \cite{PierreAuger:2021int,2019EPJWC.19703003C,Colalillo:2017uC} observation of several  anomalous events   that cannot be explained as conventional cosmic rays (CR) events.   We overview the corresponding unusual features   below in details.  A proposed answer \cite{PierreAuger:2021int} on the nature of these exotic events  is based on the observed  correlation  with the lightning events. Therefore the exotic events (EE), according to   \cite{PierreAuger:2021int} must be related to the lightning flashes, similar to the  terrestrial gamma ray flashes (TGF)  observed from satellites   above thunderstorms. Still, a complete picture and explanation of these EE   remains  obscure as many questions and puzzles yet to be answered. In particular, one of the most pressing question is as follows. Why these exotic events are so {\it rare}
 such that only less than 2 events/year had been recorded while the number of the lightning flashes are much more frequent by a large factor (at least by three orders of magnitude)?
 
 In this work we shift the question from the lightning strike per se   to the question on a possible physical source which could  ignite  and trigger a specific class of lightning flashes, which consequently may generate the EEs.  
 To be more precise,  we advocate an  idea  that the observed by AUGER the rare exotic events  could be result of a special class of lighting events  which are initiated by  the AQN dark matter particles when they   randomly hit the area of the AUGER detector  during the  thunderstorms. We will argue in this work that our proposal is consistent with  the  observations  \cite{PierreAuger:2021int,2019EPJWC.19703003C,Colalillo:2017uC}, including the    event rate and the  time duration of   EE.
 
  One should mention here that  the  AQN model   was  invented long ago without any relation  to the AUGER  observations. Rather, it was invented to 
   explain the observed  similarity between the dark matter (DM) and the visible densities in the Universe, i.e. $\Omega_{\rm DM}\sim \Omega_{\rm visible}$ without any fitting parameters, see recent brief review article on the AQN model \cite{Zhitnitsky:2021iwg}. The important feature of our proposal is that the AQN model involves no fine-tuning of parameters because this model is not devised to match the observation of EEs, but rather its properties are well studied and constrained by numerous unrelated phenomena in the previous studies for completely different purposes in a very different context in dramatically different systems,  see overview of this model in next Sec. \ref{AQN}.  
 
We  now  list the basic features of the EE observed by AUGER:\\
1. These events characterized by long lasting signals, tens  $\mu \rm s$,   and event footprints (which is defined as the total area covered by the particles emitted within this short time scale measured in tens  $\mu \rm s$) much larger, up to 200 $\rm km^2$,  than those produced by the highest energy CR;\\
 2.   In the complete data sample (recorded since 2004) 23 events with the same characteristics have been identified as exotic events
  \cite{PierreAuger:2021int};\\
3. The ``large" events are characterized by the involvement of a large number of stations (the number of triggered stations can reach  80 and more). The radius of large events spans from 4 to 8 km while the ``small" events are characterized by radius around 2-3 km. \cite{2019EPJWC.19703003C};\\
4. Large events, coined as surface detector (SD) rings in  \cite{PierreAuger:2021int}  are  often   accompanied by ``smaller" events within 1 ms, see Fig 3 in   \cite{PierreAuger:2021int}. These accompanying `` small" events are likely to be generated by a common source   but with a long time evolution (not compatible with the speed of light propagation);\\
 5. There are different types of events, the so-called SD disks which also exhibit long signals (similar to SD rings), however they are characterized by smaller amplitudes, but could be large in size, see Fig 4 in  \cite{PierreAuger:2021int};\\
\exclude{6. The energy deposited at the ground varies between $(10^{17}-10^{18})$ eV \cite{2019EPJWC.19703003C,Colalillo:2017uC}. It is in fact underestimation of the deposited energy as it does not account for the central hole in SD rings of the footprint   which is likely to be an artifact of the trigger and acquisition chain  \cite{PierreAuger:2021int}. This energy is about two orders of magnitude higher than a similar deposited energy produced by a CR shower initiated by a proton with energy $10^{19}$ eV;\\
}
6. About 70\% of the small sample of ``good events" consisting of 10 events  are correlated in time within 1 ms with lightning strikes \cite{2019EPJWC.19703003C}. The correlation remains  strong ($\sim$ 41\%) even when all EE are included, in which case the time difference between EE and strikes spans from 10 $\mu \rm s$ to 100 ms \cite{Colalillo:2017uC}.  \\
7. Some of the recorded events appear to  form the clusters entering the same zone of the array within few ms, see Fig 4 in \cite{Colalillo:2017uC}.

  In the present work we advocate a proposal that the  AQNs  are the  key ingredients   in generation of these exotic events observed by AUGER. To be more specific, the AQNs in our proposal play the  dual role. First, the AQNs may initiate and trigger a special class of  lightning strikes which are much stronger and faster than a typical lightning strike. These rare AQN-induced lightning strikes     generate the ``large" events (as defined in item 3 above), according to the proposal. It explains a strong correlation of the EEs with the thunderstorms. Secondly, the very same AQNs  in the background of electric field under thunderclouds will  emit a cluster of particles. These ``direct" emissions  from AQNs     represent the ``small" events  (as defined in item 3 above) which very often accompany  the large events. The small events will also inevitably  occur under the thunderclouds even when a  lightning strike is not ignited, and consequently, a large event cannot be generated. 
  
  One should comment here that similar unusual events have been recorded previously by Telescope Array (TA) Collaboration and coined  as the TA bursts   \cite{Abbasi:2017rvx,Okuda_2019}.  The  ``mysterious bursts"  were defined as the CR-like events  
 when at least three air showers  were recorded within 1 ms.  It has been observed that the  TA bursts are  correlated  with the lightning strikes.   Both these features are  also present in EEs observed by AUGER, see items 6 and 4 correspondingly.  We argued in \cite{Zhitnitsky:2020shd} that these 
  ``mysterious  bursts" could be a consequence of the same AQN annihilation events under the thunderstorm.  It has been also proposed in   \cite{2020JGRD..12531940B,Abbasi:2017muv,Remington:2021jxb} that  similar unusual  events can be  identified with Downward  TGFs.  We comment on  relation of the results   \cite{Abbasi:2017rvx,Okuda_2019,Zhitnitsky:2020shd,2020JGRD..12531940B,Abbasi:2017muv,Remington:2021jxb} 
 devoted to the    TA bursts and TGFs with the AUGER exotic events in terms of  the   AQN proposal
in Sect.  \ref{EE}. 

   Our presentation is organized as follows. In
subsection \ref{basics}  we overview the basics of the AQN model, while in subsection \ref{earth}   we overview the features of the AQN model which will be   relevant for the present work.   In Sec.  \ref{sec:event_rate}  we estimate the corresponding event rate. In Sec. \ref{proposal}  we describe the outcome of the    AQN   annihilation events  and argue that  the AQN may play the role of   trigger which initiates and ignites very unusual (and rare)   powerful   lightning strike. We formulate our proposal in Sect.
  \ref{EE} where we identify these rare unusual lightning flashes with AUGER exotic events. In particular, we  argue that all the unusual properties listed above can be  interpreted in terms of the AQN annihilation events propagating under the the thunderstorm. We also make few comments on  relation  between  TA bursts and TGFs with the AUGER exotic events within our  AQN proposal. 
   We  conclude with Sec. \ref{conclusion} where 
 we  suggest possible tests   which may support or refute our proposal.

 \section{The AQN   DM model }\label{AQN}
 We overview  the basics ideas of the AQN model in subsection \ref{basics},  while  in subsection 
 \ref{earth} we list    some  specific features of the AQNs  traversing the atmosphere (such as internal temperature, level  of ionization, etc).  These characteristics  will be important  for the present study  interpreting the AUGER  EEs as a special class of lightnings strikes  initiated by the AQNs when they enter     the thundercloud regions  characterized by a strong    electric field.  
 
 \subsection{The basics}\label{basics}
   The   AQN DM  model  was suggested in \cite{Zhitnitsky:2002qa}  with a single goal to 
  explain the observed  similarity between the DM and the visible densities in the Universe, i.e. $\Omega_{\rm DM}\sim \Omega_{\rm visible}$.
  This feature represents a generic property of the construction   \cite{Zhitnitsky:2002qa} as both component, the visible, and the dark are proprtional to one and the same fundamental dimensional constant of the theory, the $\Lambda_{\rm QCD}$. 
 The AQN construction in many respects is 
similar to the Witten's quark nuggets, see  \cite{Witten:1984rs,Farhi:1984qu,DeRujula:1984axn},  and    review \cite{Madsen:1998uh}. This type of DM  is ``cosmologically dark'' as a result of smallness of the parameter relevant for cosmology, which is the cross-section-to-mass ratio of the DM particles.
This numerically small ratio scales down many observable consequences of an otherwise strongly-interacting DM candidate in form of the AQN nuggets.  

The original motivation for the AQN model  can be explained as follows. 
It is commonly  assumed that the Universe 
began in a symmetric state with zero global baryonic charge 
and later (through some baryon-number-violating process, non-equilibrium dynamics, and $\cal{CP}$-violation effects, realizing the three  famous  Sakharov criteria) 
evolved into a state with a net positive baryon number.

As an 
alternative to this scenario, we advocate a model in which 
``baryogenesis'' is actually a charge-separation (rather than charge-generation) process 
in which the global baryon number of the universe remains 
zero at all times.   This  represents the key element of the AQN construction.

 In this model, the unobserved antibaryons  comprise 
dark matter being in the form of dense nuggets of antiquarks and gluons in the  color superconducting (CS) phase.  
The result of this ``charge-separation process'' are two populations of AQN carrying positive and 
negative baryon number. In other words,  the AQN may be formed of either {\it matter or antimatter}. 
However, due to the global  $\cal CP$ violating processes associated with the so-called initial misalignment angle $\theta_0$ which was present  during 
the early formation stage,  the number of nuggets and antinuggets 
  will be different.
 This difference is always an order-of-one effect irrespective of the 
parameters of the theory, the axion mass $m_a$ or the initial misalignment angle $\theta_0$.

The presence of the antimatter nuggets in the AQN  framework is an inevitable and the direct consequence of the 
    $\cal{CP}$ violating  axion field  which is present in the system during the  QCD time. As a result of this feature      the DM density, $\Omega_{\rm DM}$, and the visible    density, $\Omega_{\rm visible}$, will automatically assume the  same order of magnitude densities  $\Omega_{\rm DM}\sim \Omega_{\rm visible}$  as mentioned above.

  We refer to the original papers   \cite{Liang:2016tqc,Ge:2017ttc,Ge:2017idw,Ge:2019voa} devoted to the specific questions  related to the nugget's formation, generation of the baryon asymmetry, and  survival   pattern of the nuggets during the evolution in  early Universe with its unfriendly environment. We also refer to a recent brief review article \cite{Zhitnitsky:2021iwg} which explains a number of subtle points on the formation mechanism, see also 
  independent analysis \cite{Santillan:2020lbj} supporting the basic elements on the formation and survival pattern of the AQNs during the early stages of the evolution,  including the Cosmic Microwave Background (CMB) and  Big Bang Nucleosynthesis (BBN) epochs.   

For the present studies, however,  we take the  agnostic viewpoint, and assume that such nuggets made of {\it antimatter} are present in our Universe today irrespective to their   formation mechanism. This assumption is consistent with all presently available cosmological, astrophysical and terrestrial  constraints as long as  the average baryon charge of the nuggets is sufficiently large as we review  below.

 The strongest direct detection limit\footnote{Non-detection of etching tracks in ancient mica gives another indirect constraint on the flux of   DM nuggets with mass $M> 55$g   \cite{Jacobs:2014yca}. This constraint is based on assumption that all nuggets have the same mass, which is not the case  for the AQN model.} is  set by the IceCube Observatory's,  see Appendix A in \cite{Lawson:2019cvy}:
\be
\label{direct}
\la B \ra > 3\cdot 10^{24} ~~~[{\rm direct ~ (non)detection ~constraint]}.
\ee
The basic idea of the constraint (\ref{direct}) is that IceCube  with its surface   area $\rm \sim km^2$  has not detected any events during  its 10 years of observations. In the estimate  (\ref{direct})
it was assumed that the efficiency of  detection of a macroscopically large nugget is 100$\%$ which excludes AQNs with small baryon charges 
$\la B \ra < 3\cdot 10^{24}$ with   $\sim 3.5 \sigma$ confidence level.

Similar limits are   also obtainable 
from the   ANITA 
  and from  geothermal constraints which are also consistent with (\ref{direct}) as estimated in \cite{Gorham:2012hy}. It has been also argued in \cite{Gorham:2015rfa} that that AQNs producing a significant neutrino flux 
in the 20-50 MeV range cannot account for more than 20$\%$ of the DM 
density. However, the estimates \cite{Gorham:2015rfa} were based on assumption that the neutrino spectrum is similar to  the one which is observed in 
conventional baryon-antibaryon annihilation events, which is not the case for the AQN model when the ground state of the quark matter is in the 
colour superconducting (CS) phase, which leads to the dramatically different spectral features.  The  resulting flux computed in \cite{Lawson:2015cla} is perfectly consistent with observational constraints. 

The authors of Ref. \cite{SinghSidhu:2020cxw} considered a generic constraint for the nuggets made of antimatter (ignoring all essential  specifics of the AQN model such as quark matter  CS phase of the nugget's core). Our constraints (\ref{direct}) are consistent with their findings including the CMB and BBN, and others, except the constraints derived from    the so-called ``Human Detectors". 
As explained in \cite{Ge:2020xvf}
  the corresponding estimates of Ref. \cite{SinghSidhu:2020cxw} are   oversimplified   and do not have the same status as those derived from CMB or BBN constraints.  

 While ground based direct searches   
offer the most unambiguous channel
for the detection of the conventional DM candidates such as  Weakly Interacting Massive Particles (WIMP),  
the flux of AQNs    is inversely proportional to the nugget's mass   and 
consequently even the largest available conventional DM detectors are incapable  to exclude  (or even constrain)   the  potential mass range of the nuggets. Instead, the large area detectors which are normally designed for analyzing     the high energy cosmic rays are much better suited for our studies of the AQNs as we discuss in next section \ref{earth}.

\begin{figure}[h]
	\centering
	\captionsetup{justification=raggedright}
	\includegraphics[width=0.8\linewidth]{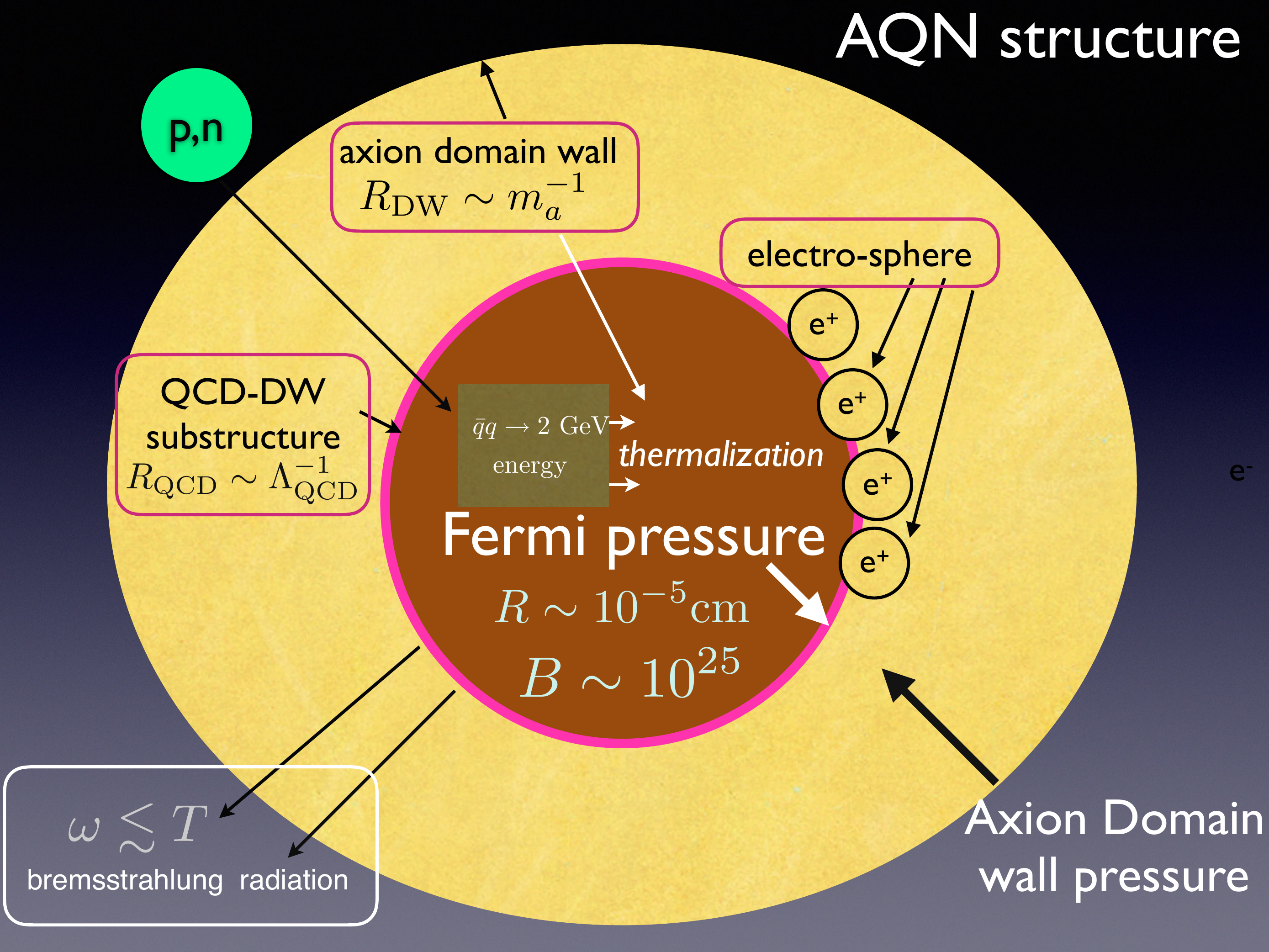}
		\caption{AQN-structure (not in scale). There are several parametrically distinct scales of the problem: $R\sim 10^{-5}$ cm represents the size of the nugget filled by quark matter with $B\sim 10^{25}$ in CS phase. Much larger scale  $R_{\rm DW}\sim m_a^{-1}$  describes the axion DW   surrounding the quark matter. The axion DW has the QCD substructure which has typical scale of order $R_{\rm QCD}\sim \rm fm$. Finally, there is always electro-sphere which represents a  very generic feature of quark nuggets, including the Witten's original construction. In case of antimatter-nuggets the electro-sphere comprises the positrons.  Its typical size strongly depends on the environment and internal temperature $T$ of the quark matter as discussed in the main text.}
\label{AQN-structure}
\end{figure}

 The absolute stability of the AQNs 
in vacuum is a result of the construction when the energy per baryon charge in   the quark-matter nuggets is smaller than in the baryons (from hadronic phase) making   up the visible portion of the Universe.
The same feature also holds for the original  {theoretical} construction \cite{Witten:1984rs,Farhi:1984qu,DeRujula:1984axn,Madsen:1998uh}. However, the difference is that, in the original model \cite{Witten:1984rs,Farhi:1984qu,DeRujula:1984axn,Madsen:1998uh},  the  quark nuggets   are assumed to be absolutely stable at zero pressure, while in the AQN model this stability is achieved by  the additional axion domain-wall pressure, see Fig. \ref{AQN-structure} with an explanation of the AQN construction and also a brief review \cite{Zhitnitsky:2021iwg} for the details. 

This difference has dramatic observational consequence- the Witten's nugget will turn a neutron star (NS) into the quark star if it hits the NS. In contrast,  a  matter type AQN   will not turn an entire star into a new quark phase because the  quark matter in the AQNs   is supported  by external axion domain wall pressure, and therefore, can be extended only to relatively small distance $\sim m_a^{-1}$,   which is much shorter  than the NS size. However,  the matter type AQNs  can be accumulated   in the cores of stars/planets during their long life times. This subject 
is of interest by itself,  but it is not the topic of the present work devoted to antimatter AQNs capable to generate the AUGER exotic events. The antimatter AQNs     will be  completely annihilated when they hit the  stars.

We conclude this brief review subsection with Table\,\ref{tab:basics} which summarizes the basic features of the AQNs. The parameter $\kappa$ in Table\,\ref{tab:basics}   is introduced to account for the fact that not all matter striking the nugget will 
annihilate and not all of the energy released by annihilation will be thermalized in the nuggets. The ratio $\Delta B/B\ll 1$ in the Table implies that only a small portion of the (anti)baryon charge hidden in form of the AQNs get annihilated during big-bang nucleosynthesis (BBN), Cosmic Microwave Background (CMB), or post-recombination epochs (including the galaxy and star formation), while the dominant portion of the baryon charge survives until the present time.  
 Independent analysis \cite{Santillan:2020lbj} and  \cite{SinghSidhu:2020cxw}    also support our original claims as cited in the Table\,\ref{tab:basics} that the anti-quark nuggets survive the BBN and CMB epochs. The large mass of  the nuggets along with their small sizes also  implies that  the direct head on AQN-AQN collisions are extremely rare events and  do not modify our estimates for $\Delta B/B\ll 1$.  
 
 Finally, one should mention here that the AQN model with the same set of parameters may explain a number of other puzzling observations 
in  dramatically different environments (Early Universe \cite{Flambaum:2018ohm,Lawson:2018qkc},  galactic  \cite{Zhitnitsky:2021wjb}, Solar corona \cite{Zhitnitsky:2017rop,Raza:2018gpb}, Earth  \cite{Zhitnitsky:2021qhj,Budker:2020mqk}) as highlighted in concluding section \ref{conclusion}.

 \begin{table*}
\captionsetup{justification=raggedright}
	\begin{tabular}{cccrcc} 
		\hline\hline
		  Property  && \begin{tabular} {@{}c@{}}{ Typical value or feature}~~~~~\end{tabular} \\\hline
		  AQN's mass~  $[M_N]$ &&         $M_N\approx 16\,g\,(B/10^{25})$     \cite{Zhitnitsky:2021iwg}     \\
		   baryon charge constraints~   $ [B]  $   &&        $ B \gtrsim 3\cdot 10^{24}  $      (\ref{direct})    \\
		   annihilation cross section~  $[\sigma]$ &&     $\sigma\approx\kappa\pi R^2\simeq 1.5\cdot 10^{-9} {\rm cm^2} \cdot  \kappa (R/2.2\cdot 10^{-5}\rm cm)^2$  ~~~~     \\
		  density of AQNs~ $[n_{\rm AQN}]$         &&          $n_{\rm AQN} \sim 0.3\cdot 10^{-25} {\rm cm^{-3}} (10^{25}/B) $   \cite{Zhitnitsky:2021iwg} \\
		  survival pattern during BBN &&       $\Delta B/B\ll 1$  \cite{Zhitnitsky:2006vt,Flambaum:2018ohm} \\
		  survival pattern during CMB &&           $\Delta B/B\ll 1$ \cite{Zhitnitsky:2006vt,Lawson:2018qkc} \\
		  survival pattern during post-recombination &&   $\Delta B/B\ll 1$ \cite{Ge:2019voa} \\\hline
	\end{tabular}
	\caption{Basic  properties of the AQNs adopted from \cite{Budker:2020mqk}.} 
	\label{tab:basics}
\end{table*}

 \subsection{When the AQNs hitting the Earth's atmosphere}\label{earth}
    For our present work, however,  the  most relevant studies  are related to the effects which  may occur when the AQNs made of antimatter hit  the atmosphere,
    the annihilation processes start and 
      a large amount of energy   will be injected to surrounding material, which may  be manifested in many different ways\footnote{\label{matter-nuggets}The AQNs made of matter will not experience any annihilation processes. Therefore, their  internal temperature always remains small. As a result,  they do not inject much energy into the space. Therefore, the matter AQNs do not  generate any  strong observable effects, in contrast with antimatter AQNs, and  will be ignored in the present studies.}.   
    For example, sufficiently large (and vary rare) AQNs with $B\gtrsim 10^{27}$ entering  the Earth's atmosphere could produce  infrasound and seismic acoustic waves     as discussed in   \cite{Budker:2020mqk,Figueroa:2021bab}  when the infrasound and seismic acoustic waves indeed have been recorded  by  dedicated instruments\footnote{\label{ELFO}A single observed event properly recorded by  the Elginfield Infrasound Array  (ELFO) which was  accompanied by correlated seismic waves is dramatically different from conventional meteor-like  events. In particular, while the event was very intense it has not been detected by a synchronized  all-sky camera network (visible frequency bands)  which ruled out a meteor source. At the same time this event is consistent  with   interpretation of the    AQN-induced    event because  the visible frequency bands must be strongly suppressed when  AQN propagates in atmosphere \cite{Budker:2020mqk}.}. 
   
   When the same AQNs enter the region under the thunderstorm the manifestation  could be much more profound as even  typical (and much more frequent)     AQNs with  $B\sim 10^{25}$ can produce very  strong observable effects.  It was precisely the goal of the recent work \cite{Zhitnitsky:2020shd} where it was argued that the AQN entering the thunderclouds   my explain  the recently observed puzzling  
    CR like events such as mysterious bursts  observed  by Telescope Array   \cite{Abbasi:2017rvx,Okuda_2019}. The main focus of   
    \cite{Zhitnitsky:2020shd} was the ``direct" emission by the AQNs in form of the highly energetic positrons accelerated by   a strong electric field, which is known to be present during the thunderstorms. 
    
    In contrast, the focus of the present work is the studies of possible ``indirect" impact   of the same AQNs when the  positrons and electrons (from ionized atmospheric molecules) could  serve as the triggers by  igniting and initiating    a large and strong lightning strike,  which consequently may generate the EEs   observed  by AUGER. 
        
      The goal  of this subsection  is to explain the basic features of the AQNs when they enter the dense regions of the atmosphere and the annihilation processes start. 
       The related  computations originally have been carried out in \cite{Forbes:2008uf}
 in application to the galactic environment with a typical density of surrounding visible baryons of order $n_{\rm galaxy}\sim 300 ~{\rm cm^{-3}}$ in the galactic center, in dramatic contrast with dense region in the Earth's atmosphere   when $n_{\rm air}\sim 10^{21} ~{\rm cm^{-3}}$. We review  these computations with few additional elements which must be implemented in case of propagation in the  Earth's atmosphere     when the density of the environment is much greater than in the galactic  environment.  

The total  surface emissivity due to the bremsstrahlung radiation from electrosphere at temperature   $T$ has been computed in \cite{Forbes:2008uf} and it is given by 
\begin{equation}
  \label{eq:P_t}
  F_{\rm{tot}} \approx 
  \frac{16}{3}
  \frac{T^4\alpha^{5/2}}{\pi}\sqrt[4]{\frac{T}{m}}\,,
\end{equation}
 where $\alpha\approx1/137$ is the fine structure constant, $m=511{\rm\,keV}$ is the mass of electron, and $T$ is the internal temperature of the AQN.  
 One should emphasize that the emission from the electrosphere is not thermal, and the spectrum is dramatically different from blackbody radiation.
     A typical internal temperature   of  the  AQNs for very dilute galactic environment can be estimated from the condition that
 the radiative output of Eq. (\ref{eq:P_t}) must balance the flux of energy onto the 
nugget 
\be
\label{eq:rad_balance}
    F_{\rm{tot}}  (4\pi R^2)
\approx \kappa\cdot  (\pi R^2) \cdot (2~ {\rm GeV})\cdot n \cdot v_{\rm AQN},  
\ee 
where $n$ represents the baryon number density of the surrounding material.  
The left hand side accounts for the total energy radiation from the  AQN's surface per unit time as given by (\ref{eq:P_t})  while  
 the right hand side  accounts for the rate of annihilation events when each successful annihilation event of a single baryon charge produces $\sim 2m_pc^2\approx 2~{\rm GeV}$ energy. In Eq.\,(\ref{eq:rad_balance}) we assume that  the nugget is characterized by the geometrical cross section $\pi R^2$ when it propagates 
in environment with local baryon density $n$ with velocity $v_{\rm AQN}\sim 10^{-3}c$.
The factor $\kappa$  accounts    for large theoretical uncertainties related to the annihilation processes 
of the (antimatter)  AQN  colliding with atmospheric molecules.

 From (\ref{eq:rad_balance})  one can estimate a typical internal nugget's temperature in the Earth atmosphere as follows:
 \be
 \label{T}
 T\sim 20 ~{\rm keV} \cdot \left(\frac{n_{\rm air}}{10^{21} ~{\rm cm^{-3}}}\right)^{\frac{4}{17}}\left(\frac{ \kappa}{0.1}\right)^{\frac{4}{17}},
 \ee 
 where  typical density of surrounding baryons is   $n_{\rm air}\simeq 30\cdot N_m\simeq 10^{21} ~{\rm cm^{-3}}$, where 
 $N_m\simeq 2.7\cdot 10^{19}  ~{\rm cm^{-3}}$ is the molecular density  in atmosphere when each molecule contains approximately 30 baryons. 
Thus, in the atmosphere  the internal nugget's temperature $T\simeq$ 20 KeV. 

It strongly depends    on unknown parameter $\kappa$. In this  work for the order of magnitude estimates we adopt  $ \kappa\simeq 0.1$, similar to our previous analysis performed in \cite{Forbes:2008uf,Budker:2020mqk}.     
 It also depends on the    altitude $z$ as the density $n_{\rm air}$ scales with altitude as $n_{\rm air}\propto \exp(- {z}/{h})$, where $h\simeq 8$ km.
In case if   environment contains high density of  ions   the parameter $\kappa$ effectively increases as the AQN  attracts more positively charged ions from surrounding material, which consequently may effectively increase the cross section and the rate of annihilation (resulting in larger value of $T$).  All these effects are very complicated at large $T$ and the corresponding discussions are well outside the scope of the present study. 
  
  \exclude{
 The  internal AQN temperature  had been estimated previously for a number of  cases.  It  may assume dramatically different values, mostly due to the huge difference in number density $n$ entering (\ref{eq:rad_balance}). In particular, for the central regions of the galaxy the temperature could assume high values of order   $ T_{\rm galaxy}\approx 5$\,eV. These hot anti-matter nuggets are very good UV emitters and could be the crucial missing component  in  explanation  of the mysterious and very puzzling observations of the  UV excess     \cite{Henry_2014,Akshaya_2018,2019MNRAS.489.1120A},   as argued in
 \cite{ Zhitnitsky:2021wjb}.
 
 When the AQN enters the   deep Earth's interior the temperature  assumes much  higher  values  on order  of  $T_{\rm rock}\approx  (100-200) $\,keV. Precisely this value of  $T$  had been used    in  \cite{Zhitnitsky:2021qhj} for explanation of the mysterious Multi-Modal Clustering CR-like  events  observed by HORIZON 10T \cite{2017EPJWC.14514001B,Beznosko:2019cI}.  The same high temperature had been used in \cite{Liang:2021rnv} for explanation of puzzling upward-going CR-like events
 with noninverted polarity as recorded by  the Antarctic Impulse Transient Antenna (\textsc{ANITA}) collaboration \cite{Gorham:2016zah,Gorham:2018ydl}. 
 }
 Important characteristic of the AQN propagating in the air   is   the number of direct collisions of the atmospheric molecules with AQN per unit time:
\be
\label{collisions}
\frac{dN_{\rm collisions}}{dt}\simeq (\pi R^2) \cdot   N_{\rm m} \cdot v_{\rm AQN} 
\simeq  10^{18} \left(\frac{N_{\rm m}}{2.7\cdot 10^{19} ~{\rm cm^{-3}}}\right)\rm s^{-1}, 
\ee   
where we use the same parameters we previously used in our estimate  (\ref{T}) for the internal temperature of the nuggets.
The dominant portion  of these collisions are the elastic scattering processes rather than successful annihilation events  suppressed by parameters $\kappa$ as discussed in the   text after  (\ref{T}).

Another  feature   we want to mention    which is relevant for our  present studies is the ionization  properties of the AQN. Ionization, as usual,   occurs in a system   as a result of  the high internal temperature $T$ as  mentioned  above. What happens with AQN when  the  internal temperature $T$  becomes  sufficiently  high  is that  a large number of positrons $\sim Q$ from the electrosphere of the anti-matter AQN get excited  but remain bound to the system. These positrons  are  very  weakly bound particles which   can easily leave the system as a result of elastic collisions with surrounding molecules. The corresponding parameter $Q$ can be estimated as follows: 
\be
  \label{Q}
Q\approx 4\pi R^2 \int^{\infty}_{0}  n(z, T)\rmd z\sim \frac{4\pi R^2}{\sqrt{2\pi\alpha}}  \left(m T\right)   \left(\frac{T}{m}\right)^{\frac{1}{4}} , ~~
  \ee
 where $n(z, T)$ is the local density of positrons at distance $z$ from the nugget's surface, which has been computed in the mean field approximation  in \cite{Forbes:2008uf} and has the following form
\begin{equation}
\label{eq:nz0}
n(z, T)=\frac{T}{2\pi\alpha}\frac{1}{(z+\bar{z})^2}, ~~ ~~\bar{z}^{-1}\approx \sqrt{2\pi\alpha}\cdot m  \cdot \left(\frac{T}{m}\right)^{\frac{1}{4}}, ~~~~  \end{equation}
where $\bar{z}$ is the integration constant is chosen to match the Boltzmann regime at sufficiently  large $z\gg \bar{z}$. Numerical 
studies ~ \cite{Forbes:2009wg}  support the approximate analytical  expression (\ref{eq:nz0}).

 Numerically, 
the number of weakly bound positrons can be estimated from (\ref{Q}) as follows:
\be
\label{Q1}
Q\approx 1.5\cdot 10^{11}   \left(\frac{T}{10~ \rm keV}\right)^{\frac{5}{4}}  \left(\frac{R}{2.25 \cdot10^{-5} \rm cm}\right)^{2}.
\ee
 These positrons from electrosphere being in the equilibrium  (when the AQN experiences a  relatively small  annihilation rate) will normally occupy very thin layer    around the  AQN's quark core as computed in \cite{Forbes:2008uf,Forbes:2009wg}. However, in our case when the AQN enters the Earth's atmosphere    a large number of non-equilibrium processes (such as generation of the shock wave  resulting from large Mach number) are expected to occur. Furthermore,    the positron's cloud  is expected to expand well beyond the  thin layer  around the core's nugget
 as a result of   high temperature $T$. A typical capture radius $R_{\rm cap}(T) \gg R$ when the positrons remain to be bound to the AQN can be estimated as 
 \be
\label{capture1}
R_{\rm cap}(T)\simeq \frac{\alpha Q(T)}{T}\sim 2~{\rm cm} \left(\frac{T}{10 \rm ~keV}\right)^{1/4},
\ee
where $Q$ is estimated by  (\ref{Q}),  (\ref{Q1}).  
The idea of this estimate is that the typical positron's excitation energy $\sim T$ must be equilibrated by the positron's binding energy $\sim \alpha Q/r$   for the particles  to remain loosely  bound to the AQN, which represents our estimate (\ref{capture1}).  

 Precisely some of these weakly bound positrons may get accelerated to the MeV energies in the background of strong electric field and mimic the CR events (termed  as the mysterious bursts) as suggested in \cite{Zhitnitsky:2020shd}
 and overviewed  in  Sect. \ref{proposal}. 
 These events will be coined as the {\it ``direct"} emissions because these positrons can be directly detected by a surface detector. 
 
 The same positrons from electrosphere can also serve as the ``seed" particles  which are always required to initiate a  lightning flash.
These non-relativistic positrons along with the electrons (which appear as a result of ionization of the atmospheric molecules)  are coined as {\it ``triggers"} in this work, because they are capable to trigger  and ignite a special class of the unique and powerful   lightning strikes to be discussed in  details in Sect. \ref{proposal} and which are accompanied by  the EEs observed by AUGER.  
 
In what follows  we   assume that, to first order, that the  finite portion of positrons  $\sim Q$   leave the system as a result of the complicated processes mentioned above, in which case    the AQN as a system   acquires a negative  electric  charge $\sim -|e|Q$  and  get   partially  ionized as a macroscopically large object of mass $M\simeq m_pB$. The ratio $eQ/M\sim 10^{-14} e/m_p$ characterizing  this object  is very tiny    such that nuggets themselves do not change momentum nor trajectory under the influence of the electric field during the  thunderstorm, and will continue to   propagate with conventional for DM particles velocity $v_{\rm AQN}\sim 10^{-3}c$.

\section{Frequency of appearance} \label{sec:event_rate} 
Here we want to estimate the total number of events which AUGER  can record within the AQN proposal during 13 years of observations (since January 2004 until 15 May 2017). 
One should emphasize from the very beginning that our estimates which follow are the order of magnitude estimates as there are many unknowns as we discuss in course of the text.
Furthermore, the observed 23  exotic events  during 13 years (less than 2 events/year)  implies that statistical fluctuations could be essential.
   The  order of magnitude estimate  presented below  is consistent with the observed rate.
   This should be considered as a highly nontrivial self- consistency check of our  proposal. 
   
   Indeed, a crucial 
   question which needs to be  answered  was formulated in the first paragraph of the Introduction which  addresses the puzzling   rareness of the exotic events. 
   In particular, an explanation   of the  EEs (in terms of the correlation   with the lighting flashes) is 
    obviously not  a satisfactory nor sufficient  one  as the dominant portion ($\sim 99.9\%$)  of the lighting events  is not accompanied by the AUGER exotic events\footnote{The authors \cite{PierreAuger:2021int,2019EPJWC.19703003C,Colalillo:2017uC}  do not provide an exact
    number of lighting flashes being recorded by AUGER during 13 years of the observations.  In our estimate  we use similar mysterious events recorded by TA collaboration when 10 bursts have been recorded during 5 years of the observations on the area where   number of lighting strikes   was recorded on the level $\sim 10^4$ which represent $0.1\%$ event ratio as quoted above.}.   
    
    The estimate presented below explicitly shows that the key missing factor in the  event rate puzzle might be  related to the rareness of the AQN dark matter events. Precisely these  events when the AQNs hit the AUGER detector area under the thunderclouds    ignite   the special class of the  extremely  rare  lightning strikes which generate EEs.    
    
     One should emphasize from the start that all relevant parameters such as the nugget's size distribution or DM density  $\rho_{\rm DM}\simeq 0.3\,{\rm  {GeV} {cm^{-3}}}$  have been used in all our previous studies not related to the AUGER exotic events and  we are not attempting to modify  any parameters of the AQN model to better fit the observed rate.  
 
 The starting point is the AQN flux which eventually determines the total number  of exotic events within AQN proposal is as follows:
 \be
\label{Phi1}
\frac{\rmd\Phi}{\rmd A\rmd\Omega}
=\frac{\Phi}{4\pi R_\oplus^2}  =  4\cdot 10^{-2}\left(\frac{10^{25}}{\langle B\rangle}\right)\rm \frac{events}{yr\cdot  km^2},
\ee
 where $R_\oplus=6371\,$km is the radius of the Earth  and  $\Phi$ is the total hit rate of AQNs on Earth \cite{Lawson:2019cvy}:
\be
\label{Phi}
\Phi
\approx \frac{2\cdot 10^7}{\rm yr}  
 \left(\frac{\rho_{\rm DM}}{0.3{\rm\,GeV\,cm^{-3}}}\right)
\left(\frac{v_{\rm AQN}}{220\kmps}\right)
\left(\frac{10^{25}}{\langle B\rangle}\right),~~~~~
\ee
where $\rho_{\rm DM}$ is the local density of DM.
The expected number ${\cal N}_{\rm EE}$ of the Exotic Events within the  AQN framework   can be  estimated as follows:
\begin{equation}
\label{eq:cal N}
{\cal N}_{\rm EE}
\approx {\cal{A}} \cdot {\cal T}\cdot {\cal F}\cdot \Delta\Omega
\frac{\rmd\Phi}{\rmd A\rmd\Omega}\,,
\end{equation}
where  ${\cal{A}} \approx 3000{\rm\,km}^2$ is the effective area of AUGER   array, ${\cal T}\approx 13\,$years is  total time of observations, 
$\cal{F}$ describes the fraction of time when the effective area ${\cal{A}} $ has been under thunderclouds, 
 $\Delta\Omega\approx2\pi$ for isotropic flux of AQNs. Factor ${\cal{F}}\simeq 0.25 \cdot 10^{-2}$  has been estimated for the TA mysterious bursts 
  \cite{Zhitnitsky:2020shd}, and we keep the same numerical value for the fraction ${\cal{F}}$ for  the order of magnitude estimates in present work as well.

Collecting all these numerical factors together we arrive to the following order of magnitude estimate:
\be
\label{eq:cal N ::num}
{\cal N}_{\rm EE}
\sim 25
\left(\frac{\rho_{\rm DM}}{0.3{\rm\,GeV\,cm^{-3}}}\right)
\left(\frac{v_{\rm AQN}}{220\kmps}\right)
\left(\frac{10^{25}}{\langle B\rangle}\right), ~~~
\ee
which is amazingly close to the   number of exotic events ${\cal N}_{\rm obs}=23$    recorded  by AUGER. In spite of the fact that   numerically Eq.  (\ref{eq:cal N ::num}) is indeed very close to the observed value, one should emphasize that this estimate should be treated as the order  of magnitude estimate, at the very best due to many uncertainties which 
enter this estimate.

First, the parameters in Eq. (\ref{eq:cal N ::num}) are in fact not precisely known. Essential parameters such as   $\rho_{\rm DM}$, $\langle v_{\rm AQN}\rangle$, and $\langle B\rangle$ only have accuracy up to order one as the local flux distribution of DM and size distribution of AQN remain unknown to date.  
  Similarly, the fraction  $\cal{F}$ may  also deviate by factor of order one   as the thunderstorm activity in these two locations could be different by factor of 2 or so.  In addition, the total number of observed EEs  could be  larger by a considerable factor than recorded by AUGER because the selection criteria being used in  \cite{PierreAuger:2021int,2019EPJWC.19703003C,Colalillo:2017uC}  are  very conservative, and may miss some small events which we call ``direct" emissions and which are not accompanied by a lightning strike as we discuss in next section \ref{proposal}. 
  
  Furthermore, the statistical significance of 23 events is obviously small, and there are many uncertainties in selecting procedure and measurements  which    may lead to misidentification of the events. The  uncertainties in particular may include such problems as 
  the lack of the signal at the center of the footprint which may or may not be   physical \cite{PierreAuger:2021int,2019EPJWC.19703003C,Colalillo:2017uC}.  In addition, some ``small" events do not pass all the quality cuts requested for a reliable reconstruction of the events because many long -lastings signals do not have the peak in the data acquisition system (DAQ), to name just a few.

In spite of  all these numerous uncertainties, we   consider the order of magnitude estimate (\ref{eq:cal N ::num}) as a highly nontrivial consistency check for our proposal  as the basic numerical factors entering  (\ref{eq:cal N ::num}) had been fixed from dramatically different physics (including solar corona heating puzzle) and  can easily deviate by large factor. 

Precisely this estimate (\ref{eq:cal N ::num}) answers (within the AQN framework) the question formulated in the first paragraph of the Introduction: why the AUGER Exotic Events are so rare? The proposed  answer is that the AQNs act as the triggers to initiate and ignite a special class of  strong lighting strikes which generate  exotic events observed by AUGER. The rareness of these exotic  events is a consequence of the rareness of the  AQN events with  very tiny DM  flux (\ref{Phi1}).

   \section{ AQNs as the triggers of the  lightning flashes}\label{proposal}

In this section we formulate the basic idea  of the proposal. We shall argue that  the AQNs  propagating  under thunderstorm   can 
initiate and ignite the special class of lighting strikes which consequently generate EEs recorded by  \cite{PierreAuger:2021int,2019EPJWC.19703003C,Colalillo:2017uC}.   Our main focus in this work is the very initial moment of the lightning events. As we review below all  lightning strikes require some sort of ``seed" particles which initiate the strike. It is normally assumed that these ``seed" particles are capable to generate a very large electron density well above the critical value   $\simeq {3\cdot 10^{8}}{\rm cm^{-3}}$  to initiate the lightning \cite{GUREVICH199979}, see also review papers  \cite{Gurevich_2001,DWYER2014147,doi:10.1063/1.1995746} devoted to detail study of the lightning strikes. It is often assumed that conventional CR can    serve as a source of required    ``seed" particles which initiate the lightning events. Our goal here  is to argue that   the AQNs are capable to provide much stronger injection of particles  to ignite and initiate a special class of the lightning events which generate EEs. We show in subsection \ref{sect:AQN-thunder} that the 
AQN will  ionize the surrounding region along its path such that the local electron density may exceed the   critical value which is required to ignite the lighting strike. We also argue that the 
number of particles   which can be emitted by AQNs and which can initiate  the AQN-induced  lightning events is much greater  in comparison with the number of ``seed" particles which could   provide  a conventional and frequent CR.  
But first, in next subsection \ref{sect:thunder} we overview the basic requirements for the lightning process to start. 
     
  \subsection{Requirements for the lightning strikes: electric field and seed particles}\label{sect:thunder} 
  In what follows we overview the basic requirements for the lighting strikes to occur. It  represents a short detour from our 
main topic. However the corresponding parameters play a key role in our arguments in following subsection \ref{sect:AQN-thunder} devoted to study of the AQNs under the thunderstorms.
 We refer to review papers \cite{Gurevich_2001,DWYER2014147,doi:10.1063/1.1995746} devoted to detail study of the lightning strikes. Here we list the basic conditions which must be satisfied for the lightning strikes to occur \cite{Gurevich_2001,DWYER2014147,doi:10.1063/1.1995746}:
  
  a) A sufficiently strong electric field ${\cal{E}} \gtrsim {\cal{E}}_c$ must   exist in thunderstorms for occurrence of runaway breakdown (RB in terminology  \cite{Gurevich_2001}) or  relativistic runaway electron avalanche (RREA in terminology \cite{DWYER2014147}). Numerical value for ${\cal{E}}_c$ is given below, see  (\ref{E}). 
  
  The basic idea here of  these processes is as follows.  When the charged particles  move in the background   of strong electric field the rate the particles  gain the energy from electric field exceeds the rate that these particles lose energy through the interaction with the air. These so-called runaway particles 
  propagate through the air produce secondary particles by hard elastic scattering with the atomic electrons, resulting in an avalanche of runaway electrons that grows exponentially with both time and distance.  The initial particles which may generate such exponential growth called seed particles, which could be electrons or positrons. 
  
  b)The spatial  scale $L_{\cal{E}}$ of a   electric field  ${\cal{E}} $ in thunderstorm must substantially exceed the scale $l_a$ for the exponential growth of the avalanche, i.e $L_{\cal{E}}\gg l_a$
as  argued  in  \cite{DWYER2014147,Gurevich_2001}.  Numerical value for $l_a$ is given below, see (\ref{l_a});

c) The presence of fast seed particles with energies exceeding the critical runaway energy 
  $E>E_c$ is absolutely necessary for initiation of the RB process.  
  The numerical value for  $ E_c$ is estimated in terms of the electric field  ${\cal{E}} \gtrsim {\cal{E}}_c$ as follows:
  \be
  \label{E_c}
  E_c\approx (mc^2)\cdot \left(\frac{{\cal{E}}_c}{2{\cal{E}}} \right).
  \ee
  In other words, the conditions a) and b) are  insufficient for RB process. It is assumed \cite{Gurevich_2001} that the high energetic  CR particles with energy $E\gtrsim (10^{15}-10^{16}) \rm ~eV$   normally generating  an extensive air shower (EAS)  
  with $(10^6-10^7)$ secondary electrons and positrons  are capable to  serve as the seed particles\footnote{It has been argued in a follow up paper \cite{PhysRevLett.110.185005} that the presence of the so-called hydrometeors in atmosphere may considerably lower the required  energy for a primary CR particle 
  such that a CR with  energy $E\gtrsim (10^{11}-10^{12})~ \rm eV$  according to \cite{PhysRevLett.110.185005} can ignite the lightning strike. In our estimates (\ref{CR-parameters}) which follow we use the original  characteristics  \cite{Gurevich_2001} for number of the secondary particles which can serve as the seed particles.}. These seed particles from CR  will  initiate and ignite the lighting strike.  It was precisely the reason to coin this  complicated process as the RB-EAS discharge mechanism. 
  
  The basic idea is that this process
  may generate the local electron density  above the critical value  \cite{GUREVICH199979}:
  \be
  \label{n_critical}
 n_{\rm critical} \simeq {3\cdot 10^{8}}{\rm cm^{-3}},
  \ee
  which  generates a highly conductive region and initiates  the lightning flash in the presence of a strong electric field. 
  
   We start by quoting the so-called critical electric field ${\cal{E}}_c$ which must exist in thunderstorms for occurrence of RB   \cite{Gurevich_2001}
    or  RREA   \cite{DWYER2014147}:
\be
\label{E}
{\cal{E}}_c= (2.16-2.84)\rm  \frac{kV}{cm} \exp\left(-\frac{z}{h}\right), ~~~ h\simeq 8 ~km.
\ee
Such strong  (and even stronger) fields are routinely observed in atmosphere using e.g. balloon measurements.   Another important characteristic  is the avalanche scale $l_a$  and the corresponding time scale $\tau_a$ for the exponential growth, which are numerically estimated as  \cite{DWYER2014147,Gurevich_2001}:
\be
\label{l_a}
l_a\simeq 10^2~ {\rm m} , ~~~~~ \tau_a\simeq \frac{l_a}{c} \sim (\rm fraction ~ of) ~\mu s.~~
\ee
The characteristic scale $l_a$ represents the minimum length scale when the exponential growth of  runaway  avalanche occurs. 
 The scale $L_{\cal{E}}\gg l_a$  essentially determines  the allowed scale of the   inhomogeneity  and   non-uniformity of a fluctuating   electric field for  the exponential growth  to hold for sufficiently long time.  
 
 The significance of the scale (\ref{l_a}) is related to the fact that precisely this scale determines the efficiency of the RB process as the intensity of the strike 
 is directly related to the exponential growth of the distribution function $f$  in time  \cite{Gurevich_2001,DWYER2014147,doi:10.1063/1.1995746}:
 \be
 \label{exp}
 f\propto \exp\left(\frac{t}{\tau_a}\right), ~~~~ N_{\rm e-fold} \equiv \left(\frac{t}{\tau_a}\right), ~~~ t\gg  \tau_a
 \ee
when number of e-foldings $N_{\rm e-fold}$ could assume very large values as parameter $\tau_a $ is measured in fraction of a microsecond. 

  While there is a consensus  on typical parameters of the electric field  $ {\cal{E}}_c$ and avalanche scales $l_a, \tau_a$ which are important characteristics for the  lightning dynamics (items ``a" and ``b" above), the physics of the  initial moment of lightning and the source of the required seed particles (item ``c" above) remains a matter of debate, and refs \cite{Gurevich_2001,DWYER2014147} represent  different views on this matter, see also
relatively  recent articles \cite{https://doi.org/10.1029/2009JA014504,doi:10.1029/2011JA016494,doi:10.1029/2011JA017431,doi:10.1029/2011JA017487} where some specific elements of existing disagreement have been explicitly formulated  and debated. 
 
 We do not wish to be involved in these ongoing debates as our main goal is not related to study of  the source, dynamics  and mechanisms related to the conventional and very frequent lightning events. We have nothing new to say on this matter. Rather the  main goal of this work is to argue that the number of particles which can be generated     by very rare AQNs traversing the thunderclouds is many orders of magnitude higher than conventional CR with energy $E\gtrsim (10^{15}-10^{16}) \rm eV$ can provide.
 
 Therefore, for our specific purposes  the disputable elements do not play any role in our studies as we simply compare the basic characteristics (such as number of injected particles, injected energy, typical time  scales, etc) between AQN-generated  and CR-generated  ``seed particles". This comparison unambiguously shows that AQN can serve as a trigger of a very special, unique  and   powerful (but very rare) event as   number of initially injected particles   is many orders of magnitude greater  than analogous  characteristics which  conventional  CR with energy  $E\gtrsim (10^{15}-10^{16})  \rm eV$ can provide.

 Therefore, it is naturally to expect that such  AQN-induced  lightning strikes should have some very distinct features  which are dramatically different from conventional   
  and very frequent CR-induced lightning strikes. We conjecture that the AUGER exotic events  \cite{PierreAuger:2021int,2019EPJWC.19703003C,Colalillo:2017uC} precisely represent these distinct features. The first argument that this idea   is at least a self-consistent proposal was presented in previous section \ref{sec:event_rate}
  where it was argued  that the computed event rate agrees well with the observed rate. Below we list the basic characteristics of a conventional  CR with energy  $E\gtrsim (10^{15}-10^{16})  \rm eV$ which according to  \cite{Gurevich_2001} can initiate a lighting strike. It  represents the {\it benchmark}  in our discussions which follow. In next subsection \ref{sect:AQN-thunder} we present some  estimates suggesting that  the dynamical characteristics of the AQNs which can serve as a trigger of a special class of the lighting events are much more powerful in comparison with these benchmark parameters   listed below. 
  
  In what follows we take for simplicity a CR with energy  of a primary particle $E\simeq  10^{15}  \rm eV$. This CR event  is approximately characterized  by the following benchmark parameters relevant for this work:
  \be
  \label{CR-parameters}
  E\simeq  10^{15}  {\rm eV}, ~~ N_{e^+e^-}\sim 10^6, ~~ \la E_{e^+e^-}\ra\sim 30~ {\rm MeV} . ~~~~~
   \ee
 \exclude{ It is assumed that such energetic CR can ignite and initiate RB process leading to a successful  lightning strike. In formula (\ref{CR-parameters}) 
  the parameter $D\simeq 15$m is the width of the ``pancake" disk, while parameter $A\simeq (150 \rm m)^2$ is the area  of this disk representing a typical EAS of a CR with energy  $E\simeq  10^{15}  \rm eV$. One should emphasize that all the parameters entering (\ref{CR-parameters}) should be understood only as the average characteristics of the EAS.  In particular, the number density of particles  $\rho$ closer to the central axis of the shower is much greater than on average. The same comment applies to the width   of the disk $D$ which is also much thinner  closer to the central axis of the shower. }Important  point here is that the EAS could  traverse the   region  with large electric field (\ref{E}), (\ref{l_a})  characterized by  distance  $L_{\cal{E}}\gg l_a$ on a $\mu$s scale which is a typical required time duration for the RB-EAS processes to start.  This process may ignite and initiate a typical lighting strike. 
    
  In next subsection \ref{sect:AQN-thunder}
 we compare    parameters (\ref{CR-parameters}) with analogous values describing  the AQN propagating under the thunderstorm. We shall see 
that the AQN  generates much more particles   in comparison with  a typical CR  benchmark parameters (\ref{CR-parameters}). Therefore, the AQN-induced lighting strikes  could be dramatically different (though much more rare) events in comparison with  typical CR-induced lighting flashes. We emphasize once again: the present subsection  represents a short detour from our main topic and  the parameters (\ref{CR-parameters}) will be used exclusively as a benchmark for illustrative purposes only with main intension of comparison them with  similar   parameters resulted from the AQN propagating  under the thunderclouds.

\subsection{The AQNs under  a thunderstorm}\label{sect:AQN-thunder}
In this subsection we study 
 the dynamics of the AQN     under influence of the pre-existing electric field characterized by parameters  (\ref{E}), (\ref{l_a}) as reviewed in  previous  subsection. There is a number of phenomena  which   dramatically affect  the initial moments of the lighting strikes
  as a result of sudden injection of a large number of particles (electrons and positrons) when the  AQN enters the region of a strong electric field.  
  First of all, some positrons from electrosphere will leave the system  as a result of this electric field. 
This effect will be  discussed  in subsection   \ref{fate-positrons} where   we estimate the number of   positrons being liberated from AQNs, their mean-free path,  and the  corresponding  energy spectrum.  Another profound AQN-induced effect  is related to the  ionization    when the electrons   are kicked out  from the atmospheric molecules and released  to the surrounding air.  We estimate the rate of the ionization,    the spectral features 
and the fate  of these short lived electrons in subsection \ref{fate-electrons}.
 
   \subsubsection{Fate of the AQN-induced liberated positrons.}\label{fate-positrons} 
   
   We start with the discussions of the positron's  fate   because the effect of the positron's liberation from electrosphere has been previously discussed in great details in relation to the so-called mysterious bursts, the CR-like events   recorded by  TA collaboration  \cite{Abbasi:2017rvx,Okuda_2019}.   To be more specific,  it has been argued in  \cite{Zhitnitsky:2020shd}  that the positrons will be liberated and accelerated  to $\sim $10 MeV energy by the electric field $\cal{E}\sim \rm kV/cm$ when the AQN propagates under the thunderstorm.  These positrons can mimic the CR events and  can be interpreted as the TA mysterious bursts.   This interpretation is consistent with the observed    intensity, timing and the frequency of appearance.
   In the present work we want to argue that the same AQN-induced  positrons
  may also play  another  role as a trigger which initiates and ignites much larger in scale a lighting strike\footnote{\label{positrons}It is interesting to note that according to \cite{DWYER2014147}  the positrons play a key role in development of the avalanche in RREA framework due to much longer mean free path  in comparison with electrons. The source of the positrons in our framework  and in  \cite{DWYER2014147}   is  completely different.}. 
   
   We begin our discussions with the following question: what happens to the weakly coupled positrons (\ref{Q1})   localized at distance   $R_{\rm cap}(T) \gg R$     when the AQN  enters the region of a strong electric field characterized by parameters (\ref{E}),    (\ref{l_a})?
 The answer suggested in  \cite{Zhitnitsky:2020shd} is as follows: a sudden appearance  of strong external electric field $\cal{E}\sim \rm kV/cm$  along the AQN's path will inject an additional energy $\Delta E$ to the positrons estimated as
\be
\label{Delta_E}
\Delta E \simeq [e{\cal{E}}\cdot R_{\rm cap}]\sim 2   ~{\rm keV}\gtrsim E_{\rm bound}  .
\ee
 This   energy injection  of order of several  keV could  liberate the weakly bound positrons  from the electrosphere.  At this moment  
 a finite portion of order $\sim Q$ of the  weakly bound positrons      will be liberated to atmosphere 
 with typical kinetic energy of order $T\sim 10$ keV.   
  These released  positrons find themselves in the background of strong electric field ${\cal{E}}$ characterized by typical length 
 scale $l_a \simeq 10^2$ m according to  (\ref{l_a}).
 This pre-existing electric field will accelerate the positrons  to $\sim 10$ MeV energies  and even higher\footnote{\label{footnote-shock}The acceleration process can be roughly divided to two different regimes: the acceleration due to the
shock waves when the positrons are accelerated from  energies $\sim 10$ keV to approximately relativistic velocities $\sim 1$ MeV, and relativistic regime when the positrons assume their high energy values $\sim 10$ MeV, see Appendix \ref{appendix-shock} for the details. One should comment here that the formation of the shock waves due to the moving AQN  is very generic feature of the framework because the Mach number $M\equiv v_{\rm AQN}/c_s\gg 1$, where $c_s$ is speed of sound in atmosphere. Therefore, a conventional Fermi's mechanism of a charged particle acceleration due to the shock wave directly applies here.}. Indeed, 
 \be
\label{E_MeV}
  E_{\rm exit}\simeq [e{\cal{E}} \cdot l_a]\sim 10 \rm ~MeV.  
\ee
We expect that a finite portion of  the released positrons ($r\sim 15\%$) is likely to be  accelerated to ultra-relativistic energies (\ref{E_MeV}), see Appendix \ref{shock-positrons} for the details. This portion of the positrons will  move   in the same direction   which is entirely determined by the  direction of the  instantaneous electric field $\vec{{\cal{E}}}$ at the moment of exit.  We termed  in the Introduction  these positrons as the {\it ``direct" } ones  because these energetic positrons have large mean free path  and   can  easily  reach  a surface detector,  producing  a  direct signal which can be recorded. 

Precisely these positrons have been identified in  \cite{Zhitnitsky:2020shd}  as a possible source of the mysterious bursts studied  by the TA collaboration \cite{Abbasi:2017rvx,Okuda_2019}.  The arguments supporting this identification are based on estimation of the  total number of the released positrons $\sim Q$ (which unambiguously predicts   the intensity of the TA signal),  the timing, the event rate and other characteristics which match the observations quite well. These ``direct" positrons could be also the source of the {\it small} events as recorded by AUGER which are characterized by relatively small size 2-3 km on the surface similar to TA mysterious bursts, see item 3 in Introduction. 

The new element here is that the very same positrons from electrosphere can also serve as the seed particles, which is absolutely required element ``c" listed at the beginning of Sect \ref{sect:thunder}.  These  initially $\sim10$ keV energy  positrons   released to atmosphere  are capable to get accelerated to relativistic energies (\ref{E_MeV}). As such these positrons   are coined as the {\it ``triggers"}  below, because they are capable   to trigger and ignite a special class of  very powerful  lightning strikes. Indeed, 
it has been known that the positrons can play an important role in development of the avalanche in RREA framework due to much longer mean free path (in comparison with electrons), see review   \cite{DWYER2014147}. Our comment here is that  the almost instantaneous injection of the large number of positrons  may dramatically 
modify the dynamics of the system during the initial moments of  the lighting strike. 
In particular,  it may  lead to the EEs   recorded by AUGER  \cite{PierreAuger:2021int,2019EPJWC.19703003C,Colalillo:2017uC}.

To put this claim in a more quantitative  way one should  compare the conventional   characteristics of the seed particles  due to the CR  as given by (\ref{CR-parameters})  and the AQN-induced positrons.  The typical number of AQN-induced particles could be as large as $Q\sim 10^{11}$ according to (\ref{Q1}) to be compared with CR-induced $N_{e^+e^-}\sim 10^6$ according to (\ref{CR-parameters}). While the AQN- induced mechanism suggests that initial energy of positrons will be of order $\sim10$ keV, the finite portion of them  could  get accelerated  to $E_{\rm exit}\sim 10~ \rm MeV$   on a time scale of   $\mu s$, see Appendix \ref{appendix-shock} for the details. This should be   compared with  CR-induced energies $\la E_{e^+e^-}\ra\sim 30~ {\rm MeV}$ according to (\ref{CR-parameters}). Therefore, the AQN-induced positrons may dramatically modify the initial moments of the lighting strikes because the  number of injected particles could be 5 orders of magnitude larger in comparison with  conventional CR-induced  parameters (\ref{CR-parameters}) with similar average energy of the particles. 

One can interpret this huge enhancement factor in terms of the  number  $N_{\rm e-fold}$ of e-foldings defined by (\ref{exp}). The claim is that the traversing AQN can almost instantaneously  inject into the system the large number of particles which is numerically equivalent to the developed RB-EAS mechanism with \be
\label{e+fold}
 N_{\rm e-fold}({e^+}) \approx \ln Q\approx 25. 
\ee
 A finite portion $r\sim 0.15$ of these positrons may even serve as seed particles as argued in
 Appendix \ref{shock-positrons}. It obviously dramatically changes the initial moments of evolution of the lightning strike 
 because the injection of these positrons occur almost instantaneously, in contrast with conventional relatively slow development of the exponential growth which would require relatively long time determined by $N_{\rm e-fold}({e^+})$.
 We identify these AQN induced events with  rare Exotic Events  recorded by AUGER  \cite{PierreAuger:2021int,2019EPJWC.19703003C,Colalillo:2017uC} as argued in next Sect. \ref{EE}.

 \subsubsection{ Fate of the AQN-induced electrons.}\label{fate-electrons}
 
 Now we turn to the analysis of the fate of the electrons which appear as a result of collisions of the AQN with atmospheric molecules
when it propagates under the thunderstorm.

 These collisions may ionize the atmospheric molecules because the AQN is a  macroscopically large     object  with   charge $Q$ estimated by (\ref{Q1}).  In this case    the direct collisions of the AQN with atmospheric molecules    can  liberate the electrons    such that  molecules become positively charged ions. Indeed,    the   AQN's   negative charge   (\ref{Q1}) implies the presence of the internal electric field ${\cal{E}_{\rm AQN}} (r)\sim eQ/r^2 $ surrounding  the nuggets.  Precisely this internal electric field due to the negatively charged quark's core may ionize the 
atmospheric molecules at the moment of collision. 

Indeed,  when the corresponding  electric potential ${\cal{U}_{\rm AQN}}(r)$  outside the nugget's core 
becomes sufficiently strong (order of several electron-volts) the   electrons from neutral molecules  could be liberated and become free electrons.
The corresponding ${\cal{U}_{\rm AQN}}(r)$ can be estimated as follows:
\be
\label{U}
{\cal{U}_{\rm AQN}}(r) \sim  \frac{\alpha Q\exp{\left(-\frac{r}{\lambda_D}\right)}}{r}  \sim T\frac{R_{\rm cap}\exp{\left(-\frac{r}{\lambda_D}\right)}}{r}.~~~~
 \ee
In equation (\ref{U})  the internal temperature $T$ is estimated  by (\ref{T}), the capture radius  $R_{\rm cap}(T)$ is given  by (\ref{capture1}), while the Debye screening length $\lambda_D$ is determined, as usual, 
by expression 
\be
\label{Debye}
\lambda_D\equiv \sqrt{\frac{T_{\rm atm}}{4\pi\alpha N_{\rm ion}} }\approx 0.4~{ \rm cm} \sqrt{\left(\frac{T_{\rm atm}}{300~ K}\right) 
\left(\frac{10^5 {\rm cm^{-3}}}{N_{\rm ion}}\right)}, ~~~~~
\ee 
where  in our numerical estimates we assume a  typical  (for RB atmospheric conditions under the thunderclouds) density of ions $N_{\rm ion}\sim 10^5 {\rm cm^{-3}}$, see e.g. \cite{Gurevich_2001}.    
From this expression is quite obvious that the strong electric potential on the level of eV scale extends to relatively large distance $r\sim$  few cm   where atmospheric molecules can get ionized as a result of a strong AQN-induced electric potential (\ref{U}). 
Indeed, one can estimate that 
\be
\label{r-ionization}
{\cal{U}_{\rm AQN}}(r)\simeq 10~ {\rm eV} ~~ {\rm at} ~~ \bar{r}\sim 3 ~{\rm cm}.
\ee
If ionization occurs closer to the nugget, the potential ${\cal{U}_{\rm AQN}}(r)$ becomes much stronger and liberated  electrons could become much more energetic. 

One should mention that the dominant portion of these liberated free electrons by the mechanism described above will obviously have  typical   energies  in the eV range which represents a typical binding energy in atoms and molecules. These free electrons (thermal electrons in terminology  \cite{Gurevich_2001})  are characterized by   very short life time of order $10^2 \rm ns$ and disappear   rather rapidly as a result of interaction with air molecules,  see e.g. \cite{Gurevich_2001}. 
The dominant portion  of the liberated   AQN-induced electrons cannot serve as seed particles, in contrast with the AQN-induced positrons discussed above  
because of very low initial energy of these electrons, see Appendix \ref{shock-electrons} with detail explanation. 
However, they could play a crucial role in initiating and triggering a unique and powerful lighting event (even though they cannot serve as the seed particles). This is because  
  the   very large number of electrons being  injected into the system almost instantaneously as estimated below     already  exceeds the critical density (\ref{n_critical})   without  necessity for the avalanche exponential growth to develop. 
The intensity of this jolt and the electron density it generates can be estimated      as follows.  

We  assume that every molecule hitting the nugget looses at least one electron by the mechanism described above. We consider it is very conservative assumption as the presence of a strong electric potential (\ref{U}) is very generic and model independent feature of the AQN framework being a direct consequence of the   internal temperature (\ref{T}) of the nuggets. With this assumption 
 we arrive to the following estimate for the number of electrons suddenly (on $\mu \rm s$ time scale) injected into the system:  
\be
\label{injected}
N({e^-})\sim \frac{dN_{\rm collisions}}{dt}  [0.1\mu \rm s]
\sim  10^{11} \left(\frac{N_{\rm m}}{2.7\cdot 10^{19} ~{\rm cm^{-3}}}\right)  ~~~~
\ee
where we used formula  (\ref{collisions}) for the rate  of the AQN collisions with atmospheric molecules, and $[0.1\mu \rm s]$ as  
a life time (attachment time) of a free thermal electron which rapidly disappears   as a result of interaction with air molecules,  see  \cite{Gurevich_2001}.
The same time interval  $[0.1\mu \rm s]$ is also   a typical time scale of the RB process as given by (\ref{l_a}).   

The large number of electrons (\ref{injected}) instantaneously  injected into the system can be interpreted 
in terms of the $N_{\rm e-fold}$ similar to our interpretation for the positrons from previous subsection: 
\be
\label{e-fold}
N_{\rm e-fold}({e^-}) \approx \ln N({e^-})\approx 25. 
\ee
One should emphasize that these electrons are low energy electrons and cannot serve as the seed particles as already mentioned. The formula   (\ref{e-fold}) is given here exclusively for the illustrative purposes to  interpret  the large number suddenly injected electrons in terms of the e-folding which is conventional way 
to describe efficiency of the  avalanche exponential growth.  Sudden injection of large number of electrons  obviously dramatically changes the initial moments of evolution of the lightning strike 
 because the injection of these electrons  occur almost instantaneously (on time scale of order $0.1\mu \rm s$ according to \ref{injected}), in contrast with conventional relatively slow development of the exponential growth which would require relatively long time determined by $N_{\rm e-fold}({e^+})$.
 
 One can represent this enormously powerful jolt of low energy electrons  in terms of the electron number density $n_e(r)$ surrounding a slowly moving nugget. 
 For an order of magnitude estimate  one can use   a simple  picture when a moving nugget during time $t$ is filled by  electrons along its path in volume of the cylinder $(\pi \bar{r}^2)\cdot  (v_{\rm AQN} t)$. One can show that  the released  electrons mostly  stay in the same location where they had been liberated  as the diffusion coefficient is small, see  Appendix \ref{diffusion} with details. Therefore, we arrive to the following estimate for the average density of the electrons surrounding the nugget:
    \be
 \label{e-density}
 \la n_e (r)\ra \sim \frac{[dN_{\rm collisions}/dt] \cdot t}{(\pi \bar{r}^2)\cdot  (v_{\rm AQN} t)}  
 \sim \frac{10^{9}}{\rm cm^3}, 
 \ee
 where $ \bar{r}\sim 3 ~{\rm cm}$ as estimated in (\ref{r-ionization}). One should emphasize that this is very conservative estimate and the local electron density could be much greater than (\ref{e-density}) as already mentioned in deriving formula (\ref{injected}).

It is instructive to compare the density  (\ref{e-density}) 
  with studies \cite{https://doi.org/10.1029/2009JA014504}   where it has been shown that the upper limit on low energy electron density resulting from runaway electron avalanches produced by a $10^{17} \rm eV$ CR shower lies in the window $n_e\in (10^4-10^7) \rm cm^{-3}$, see Fig 9 in   \cite{https://doi.org/10.1029/2009JA014504}. It is many orders of magnitude smaller than the local instantaneous density   (\ref{e-density})  produced by AQN which may ignite and initiate a very powerful (but very rare)  lightning event.  

  Significance    of  the relation  (\ref{e-density}) is  that the electron density surrounding the nugget  exceeds the critical value 
  (\ref{n_critical}).     Such high electron density dramatically modifies the polarization processes  and changes the distribution of the thunderstorm electric field.  As a result, it  generates a highly conductive region and initiates  the lightning flash \cite{GUREVICH199979,https://doi.org/10.1029/2009JA014504}. 
  The RB and RREA processes of the exponential growth of the particles are not even required in this case  as (\ref{e-density}) is already sufficient to start the lightning. 
 This jolt of electrons  (\ref{injected}) producing enormous electron density  (\ref{e-density}) obviously   must dramatically change the initial moments of evolution   of the lightning strike.  We identify  the   unusual and  powerful 
  AQN-induced  lighting strikes with exotic events being  recorded by AUGER.

 \section{AUGER Exotic Events as the AQN-induced  phenomena }\label{EE}
 Our main goal here is to argue that the very unusual features of the Exotic Events as listed by items 1-7 in the Introduction could be explained within the AQN framework when these events are  induced by the AQN propagating under thunderclouds. As explained in previous section we identify the Exotic Events with  rare events of the AQNs propagating under the thunderclouds, i.e.
 \be
 \label{identification}
 {\rm AUGER ~Exotic ~Events} ~\equiv ~{\rm AQN-induced ~ events }. ~~~~~
 \ee
 
 The corresponding annihilation events due to interaction of the anti-matter nuggets with atmospheric molecules generate enormous injection of the positrons and electrons into the surrounding region. The corresponding effects  can be expressed in terms of the  e-fold numbers  $ N_{\rm e-fold} $ given by   (\ref{e+fold}) and  (\ref{e-fold}) correspondingly.  Furthermore, the same annihilation events generate enormous electron density along the AQN path as given by  (\ref{e-density}).  We coin these injected particles as the {\it triggers} because they can trigger, ignite and initiate a very powerful (and very distinct)  lighting strike if other conditions, such as presence of a strong electric field along the AQN path, are met.  If the AQN-induced lighting strike indeed does occur, it must be very distinct from conventional and typical flashes as the initial moments 
 of the AQN-induced strike are  dramatically different from   typical and much more frequent lightning strikes induced by any  conventional possible sources such as  cosmic rays.  
 
 \subsection{ AQN proposal  (\ref{identification}) confronts the  observations}\label{sec:confronts} 
 
 The first argument supporting our identification (\ref{identification}) is the estimation of the frequency of appearance of these events  (item {\bf 2}) which is consistent with our estimate presented in Sect. \ref{sec:event_rate}.
 One should emphasize that this is a very nontrivial consistency check (or ``mysterious coincidence" for sceptics) as the event rate ${\cal N}_{\rm EE}$ as given by (\ref {eq:cal N ::num}) is expressed in terms of apparently unrelated parameters such as  the DM density. However, there are much more   arguments which seemingly  support this unorthodox proposal. 
  
  From our description of the physics related to the AQN propagating under the thunderclouds  it is quite obvious that some events may ignite the lighting strikes, and some cannot, depending on other features of the   thunderstorm at the moment   the AQN hitting  the area. It explains the item {\bf 6} from the list that not all the EE are 100\% correlated with the lighting strikes.
  
  The emergence of the cluster-like events  representing the item {\bf 7} is also very generic feature of the identification (\ref{identification}). Indeed, the AQN traversing the thunderclouds propagates just $[v_{\rm AQN}\cdot   \rm ms]\sim 3\cdot 10^2 \rm m$ which is much shorter than 
 a typical scale of a thunderstorm cloud system.  This implies that  the  AQN remains in  the  same area  where a lightning event is  generated. Therefore, it is not a surprise that the cluster-like events separated by a [ms] time interval are localized in the same zone of the array.  
  
 The ``large" events  described   in item {\bf 3} are identified with events   when the AQN  successfully initiates the lighting strike which must be 
 dramatically different from  typical and much more frequent lightning strikes.  Indeed, 
 in this case few first  lighting``steps"  are initiated by  enormous initial injection of the particles characterized by  (\ref{e+fold}),  (\ref{e-fold})  which are equivalent to well-developed phase of a conventional typical  strike.  The enormous electron density   exceeding  the critical value (\ref{e-density}) may also play a role of igniting the rare and very unique strikes.   The EE could be  a consequence of these initial very powerful ``steps". 
 The enormous scale of the ``large" events is determined by the energy of the lighting itself, not by the annihilation energy released by the AQNs.
 
 The ``small" events also mentioned in item {\bf 3}, on other hand,  could be   related to the energetic positrons directly emitted by AQNs, similar to our interpretation \cite{Zhitnitsky:2020shd} of the TA bursts, see next subsection  \ref{TA-bursts} with more comments on this possible connection. 
 
 To avoid confusion with the interpretation one should emphasize that the electric field in a thunderstorm area  is a strongly (temporal and spatial) fluctuating field. Therefore, the direction of the current (or the leader) of a large developed lightning event in general does not coincide with direction of the instant fluctuating electric field, which  determines the direction of the energetic positrons emitted by AQNs \cite{Zhitnitsky:2020shd}. Therefore, the ``large" and ``small" events are not necessary recorded at the same instant  by the surface detector as the electric field generating the ``small" events is likely to  point in a  different direction at the moment of emission.  To rephrase it, an AQN plays the dual role when it propagates in  the thunderclouds:  it emits the very energetic  positrons (``small" events), and it also triggers the lightning (``large" events).  The direction of current (and hot conducting leader) in large event and direction of the bunch of relativistic particles in small event are    orientated randomly   as they are induced by two different electric fields in two different regions of the thundercloud at slightly different instances separated by $\sim ms$ time interval.

Nevertheless, one should expect some sort of correlation    between ``large" and ``small" events. Such correlation  indeed has been  recorded as   described   in item {\bf 4}. This correlation  within [ms] time interval finds its natural explanation in this proposal  because both phenomena are essentially  originated from the same AQN within the same region of the thunderstorm activity. It could be enormous variation in scales for ``large" events in intensity and in size of the footprints as these are related to the few initial ``steps" of the lighting flashes.  It    explains the item {\bf 5}. It should be contrasted with ``small" events which are characterized by the internal physics of the AQNs and should not demonstrate enormous variations in shapes or   scales of the footprints.    
 
 In our proposal (\ref{identification}) the time duration of a ``large" signal is determined by the time scale of the few initial  lighting ``steps" 
 which are triggered and  initiated by  enormous   injection of the particles characterized by  (\ref{e+fold}),  (\ref{e-fold}) accompanied by  enormous electron density  along the AQN path, exceeding  the critical value (\ref{e-density}).  As the inter-step intervals last some tens of $\mu$s, the same time scale should characterize the EE which is consistent with observations listed in  the item {\bf 1}. 
 
 Relatively long time duration of a ``small" signal on the level $(4-7) \mu \rm s$ is determined by   features  of the bunch of emitted
 positrons in the background of the electric field, similar to our discussions  in  \cite{Zhitnitsky:2020shd} in application to the mysterious bursts recorded by TA. Typical time scale for ``small" events should be  similar in scale (but somewhat smaller) than a  time   variation $\tau_{\cal{E}}$ of the of the electric field 
 ${\cal{E}}$ in thunderclouds. This is  because the acceleration  of the positrons in this framework is due to this instantaneous field ${\cal{E}}$. The electric field ${\cal{E}}$ flips its direction on the time scale  $\tau_{\cal{E}}\sim 20 \mu$s \cite{DWYER2014147}, which is consistent with observed time scale  $(4-7) \mu \rm s$
 of ``small" events.

\subsection{Relation to TA bursts   and downward TGF}\label{TA-bursts}
In this subsection we want to make few remarks on possible relation between the AQN-induced events
which is the topic of the present work and ``mysterious bursts"   \cite{Abbasi:2017rvx,Okuda_2019} and  TGFs
 \cite{2020JGRD..12531940B,Abbasi:2017muv,Remington:2021jxb}  as observed by Telescope Array  Collaboration. We want to argue that
these unique,      very rare, and unusual events represent different manifestations of the same phenomena when   the AQNs enters the thunderstorm clouds and    initiate   very special, fast, powerful  and unique lighting events. 

The TA ``mysterious bursts"   are defined as the events when at least three air shower triggers were recorded within 1 ms,  which would be a highly unlikely occurrence for three consecutive conventional CR hits  in the same area within a radius of approximately several kilometres.  
Therefore, this clustering feature   of the  ``mysterious bursts"   is hard to interpret in terms of the conventional CR events. 
  This feature in all respects is very similar to item {\bf 7} as discussed in previous subsection in connection with Exotic Events    recorded by AUGER. Another distinct feature of the TA  ``mysterious bursts" is the correlation with the lighting flashes, which is also very similar to item {\bf 6} from the previous subsection devoted to EE.  

One should also note that the ``small" events recorded by AUGER with typical size of the footprints  $\sim (2-3)$ km (as listed in item {\bf 3}) are very   similar  to the TA ``mysterious bursts"   with the same  size scales of the footprints as mentioned in the previous subsection \ref{sec:confronts}. According to our interpretation these signals in both cases is a result of the same direct emission of the positrons by AQNs
traversing  the thundercloud region.
Therefore, they must be  characterized  by the same features. In particular,  these events should have  well-localized footprints, which should not    vary much  from event to event, in contrast  with ``large" events as discussed in the previous subsection \ref{sec:confronts}. It is hard to say why TA collaboration have not recorded the  signals similar to the ``large" energetic events similar to EEs.  It could be that TA collaboration has much lower photon detection efficiency as suggested in \cite{2019EPJWC.19703003C}.

Further to the point on similarity between the two phenomena.  The event rate  of the TA  ``mysterious bursts"  as estimated in  \cite{Zhitnitsky:2020shd} is consistent with frequency of appearance for the EE discussed in Sect. \ref{sec:event_rate}. Both estimates are based on one and the same formula (\ref{eq:cal N}) with identical hit rate determined by the DM flux $\Phi$. The consistency between both event rates is very encouraging sign that both phenomena are indeed related to the same physics of the DM nuggets. 

This extremely   low event rate   is a very puzzling feature in both cases  if one regards    these events as a mere    consequence of the lighting strikes because  such a simplified view does not address the crucial    element of the puzzling feature  which is  their extreme rareness in comparison with  much more frequent  conventional lighting   strikes\footnote{\label{rareness}The exact number of lightning strikes has not been recorded by AUGER. However, in case of TA mysterious events this ratio is known: during 5 years  of observations the TA Collaboration recorded just 10 ``mysterious events" to be compared with  $\sim 10^4$ conventional lighting strikes   during  the  same period of  time recorded on the same area. This implies that $99.9\%$ of conventional lightning strikes do not generate EE observed by AUGER.}.

We now turn to the possible  relation between   TGFs
 \cite{2020JGRD..12531940B,Abbasi:2017muv,Remington:2021jxb}  as observed by TA  Collaboration and EE as recorded by AUGER, which is the topic of the present work. 
 TGFs are bursts of gamma-rays initiated in the EarthÕs atmosphere, first reported in 1994 by a satellite. Since then, a number of observations have shown that satellite-detected TGFs are produced by   upward intra-cloud  flashes. Similar TGFs have been also observed by TA collaboration  \cite{2020JGRD..12531940B,Abbasi:2017muv,Remington:2021jxb} which are identified as the downward   breakdown that occurs at the beginning of cloud to ground flashes.

 The relation between TA  ``mysterious bursts"  and TGFs has been already stated  in 
 \cite{2020JGRD..12531940B,Abbasi:2017muv,Remington:2021jxb}. Therefore, our identification  of the TA bursts with the AQN-induced events  automatically implies (within the same framework) that the TGFs recorded by TA must be  also related to the same unique and very rare lightning strikes  triggered by the AQNs traversing  the thundercloud region.   
 
 The only additional comment which requires a clarification here is that the TGFs are originated from special type of flashes which are very rare and very unique, as emphasized above. The corresponding features are not shared by conventional and much more  frequent lighting strikes. What makes  these rare lightning strikes   to become so special? Our  proposal is that   an  instantaneous  injection of large number of positrons and electrons (at the initial stage of the strikes) as represented in terms of the    e-foldings  (\ref{e+fold}) and  (\ref{e-fold}) being accompanied by  enormous electron density   exceeding  the critical value (\ref{e-density}) may trigger the lighting strikes with such unusual features. Corresponding supporting arguments (suggesting that the AQN-induced lightning strikes must be dramatically different from conventional and much more frequent flashes) were presented in previous subsection \ref{sec:confronts}. 

We conclude this section with the following remark. As we already mentioned 
 the  main focus in this work is the very initial moment of the lightning events. 
 It is not the goal of the present work to  discuss the evolution (and complicated physics) during the later stages of the  strikes triggered and initiated by enormous jolt of the AQN-induced particles.  Such studies are  obviously well beyond  the scope of the present work. Our goal here is fundamentally different: 
 we want to point out that the evolution of the lighting events (including RR \cite{Gurevich_2001} or RREA \cite{DWYER2014147,doi:10.1063/1.1995746} processes) could be dramatically   modified as a result of an  initial    injection of large number of positrons and electrons  as given by   (\ref{e+fold}) and  (\ref{e-fold}) when AQNs traversing the area under thunderclouds.  The  generation of an   enormous electron density   exceeding  the critical value (\ref{e-density}) along the AQN path may also drastically alter  the conventional dynamics of the lighting events.  We propose  that   this    injection of  particles      ignites and initiates a special and unique class of the lightning events which
 are identified with the AUGER Exotic Events, according to   (\ref{identification}).

\section{Conclusion and future development}\label{conclusion}
Our conclusion is divided into three  different  subsections: first  of all, in Sect. \ref{sect:results} we   list the basic claims of this work,  in Sect. \ref{sect:tests} we describe  several   tests  to support or refute the proposal  (\ref{identification}).
  Finally, in Sect.\ref{sect:other} we list a number of other instruments which may 
 detect  the excess of   emission due to the AQN annihilation events in dramatically different environments (Early Universe, galactic scale, solar corona, Earth's atmosphere). 

\subsection{Basic Claims}\label{sect:results}
Our basic results can be summarized as follows.  We proposed that the initial    injection of large number of positrons and electrons  as estimated by   (\ref{e+fold}) and  (\ref{e-fold}) may  trigger and dramatically modify the lightning flashes  when AQNs traversing the area under thunderclouds. The electron density   exceeding  the critical value (\ref{e-density}) along the AQN path may also ignite and initiate a  unique and special class of the lighting flashes. According to our proposal this  unique class of the lightning events  recorded by 
 AUGER   are identified with the AQN induced events 
  (\ref{identification}). The event rate as estimated by (\ref{eq:cal N ::num}) is consistent with the   number of exotic events ${\cal N}_{\rm obs}=23$    recorded  by AUGER. Therefore,  the very puzzling feature of  extreme rareness of the EE in comparison with much more frequent conventional lighting strikes finds its natural resolution within the AQN framework as it is determined by very tiny DM flux (\ref{Phi1}), see also footnote \ref{rareness}. 

\subsection{Possible  future  tests}\label{sect:tests}

In this subsection we want to discuss possible tests which can support or refute our proposal  on identification AUGER Exotic Events with  the AQN-induced events (\ref{identification}). Some of the ideas  have been already mentioned in our  paper \cite{Zhitnitsky:2020shd}
where we argued that the TA  ``mysterious bursts"  is one of the manifestations of the AQN-induced events. As these bursts are identified with   ``small" AUGER Exotic Events, the same tests also apply here.   We shall not repeat these suggestions in the present work by referring to the original paper   \cite{Zhitnitsky:2020shd}.

We think that the most unambiguous  test  which can  discriminate  our proposal (\ref{identification})
 from any other suggestions is the study of the radio pulses
(as suggested in \cite{Liang:2021wjx})  which always accompany the AQNs traversing under the thunderclouds.  The corresponding radio pulses are dramatically different from conventional radio signals during thunderstorms and from CR events.

 Indeed,  it has been known for many years that  the   lightning flashes are always accompanied  by the  radio emission.   This is a very generic feature of the  the  lightning discharges and it is well documented, see  e.g. \cite{GUREVICH2003228}  with large number of references on  the original results.  We want to mention some of these well known  features in the  text below to compare with the similar properties   of the radio pulses   accompanying  the AQN-induced events as computed in \cite{Liang:2021wjx}. 
 
  It is known that the     lightning discharges are  characterized by a very large number of radio pulses which normally last in total for about 1 second.  Each pulse is characterized by full width $(0.2-0.3)\mu {\rm s}$. The   electric field strength  of these pulses   could be as large as  $ |\mathbf{E}|\sim10^3  \rm ~mV/m$
  at distance about 10 km. Most of the pulses, though,  show  the strength of the electric field in the   
$|\mathbf{E}|\sim (100-200) \rm ~mV/m$ range.  Another important  feature of the radio emission the     lightning discharges is as follows.  The gaps between pulses are in the range $ (10-10^2) \mu {\rm s}$. Therefore, total number of pulses could be very large $\gtrsim 10^3$ during a single lightning event.  Finally, the key element of the the radio emission during  the     lightning discharges is its  typical frequency band.  The radiation is strongly peaked in few MHz bands, while it completely diminishes for $\nu\gtrsim 10 \rm ~MHz$.

  These features must be contrasted with  the AQN-induced radio pulses studied in \cite{Liang:2021wjx}. 
 Indeed, the number of clustered  radio pulses associated with the AUGER Exotic Events (and also with TA mysterious bursts)  must be very few
 to correspond to several  events which often accompany the ``large" signal, see item {\bf 4} from the list.
  It should be    contrasted with $\sim 10^3$ in case of the conventional 
lightning-induced radio pulses.  

The most important distinct feature which discriminates the different sources of the emission is that the frequency bands of the radiation are dramatically different  for EE in comparison with conventional lightning flashes, which  are strongly peaked in few MHz bands, as mentioned above.
The  AQN-induced radio pulse is characterized by the flat spectrum with $\nu\lesssim 200 \rm ~MHz$ according to  \cite{Liang:2021wjx}. The basic reason for  this dramatic difference is that the electric current responsible for a typical  lightning flash is dominated  by the particles with $\gamma\sim1 $ while for the AQN-induced case the injection and acceleration of the large number of  positrons is  characterized by $\gamma\sim 20$, where $\gamma=E/m$ is conventional relativistic factor.
  This difference is translated into dramatic modification of the frequency bands     according to  \cite{Liang:2021wjx}. 
  
 Therefore, one should not expect any difficulties  to discriminate  the AQN-induced radio pulse (which always accompanies the EE) from a conventional radio emission during the thunderstorm. The corresponding studies of the radio pulses characterized by the flat spectrum with $\nu\lesssim 200 \rm ~MHz$ (and  which must be   synchronized with every event from the cluster of the AUGER Exotic Events) could  support, substantiate  or refute this proposal.

 \subsection{Possible  future  tests with other instruments}\label{sect:other}
 There are several  other unusual and mysterious  observations of the  CR-like events which might be related to the AQN propagating in atmosphere.     It includes, along with previously mentioned TA bursts,   the   \textsc{ANITA}  observation    \cite{Gorham:2016zah,Gorham:2018ydl}  of two anomalous events    with noninverted polarity which can be explained within AQN framework \cite{Liang:2021rnv}. It also includes    the  Multi-Modal Clustering  Events observed by HORIZON 10T \cite{2017EPJWC.14514001B,Beznosko:2019cI} which impossible to understand  in terms of the CR events, but  which could be  interpreted in terms of  the  AQN annihilation events in atmosphere   as argued in  \cite{Zhitnitsky:2021qhj}. Similar mysterious  CR-like  events can also manifest themselves in form of the acoustic and seismic signals, and could be in principle recorded if dedicated instruments are present  in the same area where CR detectors are located. In this case the synchronization between different types of instruments could play a vital role in the discovery of the DM.
 In fact, in \cite{Budker:2020mqk}  it has been suggested to use distributed acoustic sensing (DAS) to search 
for a signal generated by   an  AQN propagating in the Earth's atmosphere.   It is interesting to note that   a mysterious seismic event 
indeed has been recorded   in infrasound frequency band by Elginfield Infrasound Array  (ELFO). It    has been interpreted in  \cite{Budker:2020mqk}  as the AQN-induced phenomenon, see footnote \ref{ELFO} with details.  

Our original comment here is that the acoustic signal generated by   an  AQN propagating in the Earth's atmosphere under the thunderstorm must be very different from conventional acoustic waves which normally accompany the lightning strikes. In the AQN proposal  the corresponding acoustic pulse must be synchronized with very initial moment of the lightning events  which are triggered by enormous injection of the particles characterized  by (\ref{e+fold}), (\ref{e-fold}) and  (\ref{e-density}).
    
     The presence of the {\it antimatter} nuggets in the system implies that there will be  annihilation events leading to  large number of observable effects on different scales: from Early Universe to the galactic scales to the terrestrial rare events. In fact, there are many hints suggesting that such annihilation events may  indeed  took place in early Universe as well as they are happening now in present epoch. In particular, the  AQNs might be responsible for a resolution of   the  ``Primordial Lithium Puzzle" \cite{Flambaum:2018ohm}
 during  BBN epoch. The AQNs  may also  alleviate the tension between standard model cosmology and the recent EDGES observation of a stronger than anticipated 21 cm absorption feature as argued in \cite{Lawson:2018qkc}.   The AQNs might be also responsible for famous   long standing problem of  the ``Solar Corona Mystery"
  \cite{Zhitnitsky:2017rop,Raza:2018gpb} when the   so-called ``nanoflares" conjectured by Parker long ago \cite{Parker} are   identified with the  annihilation events in the AQN framework. 
  
 Another   very promising alternative path to search for  the AQN annihilation events is to study the   excess of the radiation in the central regions of the galaxy where DM and visible matter densities are relatively high. In particular,   the  well-established  excess of the  diffuse UV emission which cannot be explained by conventional astrophysical sources  has  indeed been recorded in recent studies \cite{Henry_2014,Akshaya_2018}.   As argued in
    \cite{Zhitnitsky:2021wjb} this puzzling diffuse UV emission can be naturally understood  within the AQN framework. One should emphasize that the corresponding estimates in dramatically  different environment   were based on the same  basic parameters of the AQN model, being used in the present work.
     Therefore, the  corresponding estimates demonstrate at least the self-consistency  of the entire framework.

      If  our  interpretation of the Exotic Events recorded by AUGER  advocated in the present work is confirmed by future studies (e.g. by analyzing  the synchronized radio pulses or acoustic signals using  DAS)
 it would represent  a strong     argument  suggesting that  the resolution of   two long standing puzzles in cosmology -- the nature of the DM   and the   matter-antimatter asymmetry of  our Universe--  are  intimately  linked. The corresponding deep connection is  automatically implemented 
 in the AQN framework by its construction.

     \section*{Acknowledgements}
     This research was supported in part by the Natural Sciences and Engineering
Research Council of Canada.

\appendix
\section{Shock Acceleration and Diffusion: application   to the AQN propagating in atmosphere.}\label{appendix-shock}
In this Appendix we overview the well known Fermi's mechanism of a particle acceleration by  a  shock wave as well as diffusion features of the emitted particles.

We apply the generic properties of Fermi's mechanism  to the specific  conditions which are realized for the AQN traversing in atmosphere under thunderclouds.
While the mechanism is very generic in principle and widely used in astrophysics, e.g. in explanation of the cosmic rays acceleration  by the shock in supernova remnants, the details are very different with dramatically different consequences.  We mostly follow the textbook \cite{padmanabhan_2001} in our formal presentation in subsection \ref{shock-basics}, while in subsections \ref{shock-positrons} 
 and \ref{shock-electrons}  we give 
 our  numerical estimates  for  the cases of the AQN-induced positrons and the AQN-induced electrons  correspondingly. Finally, in subsection \ref{diffusion}
 we make simple estimations for the diffusion coefficient to justify our estimate (\ref{e-density}) given in the main body of the text for the 
 electron's density surrounding the AQN traversing in atmosphere. 
 
\subsection{Shock Acceleration. The basics}\label{shock-basics}
As we mentioned in the main text   the formation of the shock waves due to the moving AQN  is very generic feature of the system because the Mach number is very large, $M\equiv v_{\rm AQN}/c_s\gg 1$. Therefore, a conventional Fermi's mechanism of a charged particle acceleration due to the shock  as reviewed below  directly applies here. 
In what follows we identify the velocity of the shock front with the AQN velocity $v_{\rm AQN}\sim 10^{-3}c$. 

\exclude{
We start with the formula describing the energy gain $\Delta E$ of a charged particle of energy $E$ crossing the shock front
\be
\label{shock-1}
\langle\frac{\Delta E}{E}\rangle =\frac{2}{3}\frac{V}{c}, ~~~~ V\sim v_{\rm AQN},
\ee
where averaging over all   incident angles with respect to direction of the shock front moving with velocity $\sim V$  has been performed.  
}

We start by  introducing parameter $\tau$ which describes  the mean time between the two consecutive  occurrences of the process (crossing the shock)
such that $\tau^{-1}$ describes the rate of occurrences. We also introduce factor  $\beta>1$ which describes the relative energy gain for each successful
 occurrence. We also need to introduce the time scale $T_{\rm out}$ characterizing the escape process such that $P(t)=\exp{(-t/T_{\rm out})}$ describes a probability 
 for a particle to remain  in the accelerating region for time $t$. Then $P_0=\exp{(-\tau/T_{\rm out})}$ is the probability that the particle remains within the accelerating region after a single occurrence. The number  of particles  $n(t)$  that remain  in the accelerating region   is given by 
 \be
 \label{n_t}
 n(t) \approx n_0 \exp{\left(-\frac{t}{T_{\rm out}}\right)}, ~~~~~~\frac{dn}{dt}\approx -n\left(\frac{1}{T_{\rm out}}\right).    ~~~ 
 \ee
 The mean number of occurrences in time $t$ is $t/\tau$. Therefore, the typical energy of these particles in time $t$ is given by
 \be
 \label{E_t}
 E(t)\approx E_0\beta^{\frac{t}{\tau}}=E_0\exp\left(\frac{t\cdot \ln\beta}{\tau}\right) 
 \ee
 which can be inversed to express $t(E)$ as follows
 \be
 \label{t_E}
  t(E)=\tau\left[\frac{\ln (E/E_0)}{\ln\beta}\right], ~~~ \frac{dt}{dE}=\frac{\tau}{E\cdot \ln \beta}.
 \ee
 These formulae allow us to change  variables: instead of $t$ entering (\ref{n_t}) one can use  energy $E$ using Eq.  (\ref{t_E}) relating these two variables.
  Therefore,  the number of particles $n(E)$ as a function of  energy $E$
 can be expressed as follows:
 \be
 \label{n_E}
 dn(E)\propto E^{-p} dE, ~~~ p=1+\frac{\tau}{T_{\rm out}\cdot \ln\beta}=1-\frac{\ln P_0}{\ln \beta},~~~
 \ee
 where we expressed $ {\tau}/{T_{\rm out}}\equiv -\ln P_0$ as defined above. 
 The result (\ref{n_E})  implies a power-law spectrum with a specific index $(-p)$.
 
 Computation of the parameters $P_0$ and $\beta$ is a hard dynamical problem which requires a detail studies of the Boltzmann equations. 
 In a simplified setting one can argue that $\ln \beta \approx -\ln P_0 $ such that specific index $p\approx 2$, and  $dn(E)\propto E^{-2} dE$,
 which is indeed close to the observed average value $p\approx 2.7$ in very extended  range of CR energies.

 In the next subsections \ref{shock-positrons} and \ref{shock-electrons} we apply the Fermi's mechanism overviewed above to   the AQN traversing the atmosphere.  We also    give an  order of magnitude estimates for the relevant parameters.

 \subsection{Shock Acceleration: application to the    positrons}\label{shock-positrons}

As we discussed in subsection \ref{fate-positrons} the typical energy of the AQN-induced    positrons is order of $T\sim 10$ keV, which implies that their typical velocity $v_{e^+}\sim \sqrt{2T/m}\sim 0.2 c$. The mean-free path $\lambda_{e^+}$  and typical time scale $\Delta t_{e^+}$ between the consecutive elastic collisions with atmospheric molecules for such positrons can be estimated as follows. The Coulomb cross section   is
\be
\label{cross-section}
\sigma=\frac{1}{4}\left(\frac{r_e}{v^2\gamma}\right)^2\frac{1}{(\sin \frac{\theta}{2})^4}, ~~ r_e\equiv\frac{\alpha}{m}\approx 2.8\cdot 10^{-13}\rm cm~~~~~~
\ee
Using our   parameter for $n_{\rm air}\sim 10^{21} \rm cm^{-3}$ as in the main body of the text in (\ref{T}) we arrive to the following typical parameters   
  for mean-free path $\lambda_{e^+}$  and   time scale $\Delta t_{e^+}$:
  \be
  \label{e+}
  \lambda_{e^+}\approx (\sigma n_{\rm air})^{-1}\sim {\rm 20~ cm}, ~~ \Delta t_{e^+}\approx \frac{ \lambda_{e^+}}{v_{e^+}}\sim 3 ~\rm ns,~~~~~
  \ee
  where for the numerical estimates we use $\la\sin^2 \frac{\theta}{2}\ra\simeq 1/2$. Now we can estimate  the mean time $\tau_{e^+}$ between the two consecutive  occurrences of the process as follows
  \be
  \label{tau+}
  \tau_{e^+}\sim \frac{ \lambda_{e^+}}{v_{\rm AQN}}\sim 0.7 \mu \rm s,
  \ee
  where we assume that the shock is moving with velocity $v_{\rm AQN}\sim 10^{-3} c$, while a particle is localized at distance   $\sim \lambda_{e^+}$ 
  from the front after the previous occurrence. Comparing the time scales (\ref{e+}) and (\ref{tau+}) one can infer that positrons elastically scatter $\sim 10^2$ times before the next occurrence (crossing the shock) takes place. 
  
  Our next comment is related to time scale $T_{\rm out}$ characterizing the escape process as expressed by (\ref{n_t}). The key comment here is that 
  the positrons obviously leave the accelerating region with typical time scale $T_{\rm out}$. However, in contrast with the application to the CR, the AQN produces new positrons as it propagates in the external electric field $\cal{E}$. Therefore, the basic equation (\ref{n_t}) is dramatically modified. 
  
  The point is that the positrons which leave the system will be replaced by new positrons from the electrosphere as the annihilation processes continue with the rate  (\ref{collisions}) such that the temperature remains the same as given by (\ref{T}). The equilibration implies that    number of weakly bound positrons
  (\ref{Q1}) also remains the same. This important additional ingredient in the system implies that the equation (\ref{n_t}) is modified as follows:
  \be
 \label{production}
 \frac{dn}{dt}\approx -n\left(\frac{1}{T_{\rm out}}-\frac{1}{T_{\rm in}}\right),    
 \ee
  where the  time scales $T_{\rm out}$ characterizing the escape process and $T_{\rm in}$ characterizing the rate of newly produced positrons are approximately the same as a result of equilibration, i.e. $T_{\rm out}\approx T_{\rm in} $.  This modification also affects the spectral properties of the accelerated positrons,
  which assumes the form: 
   \be
 \label{n_positrons}
 dn(E)\propto E^{-1} dE,  
 \ee
  where we neglected the second term for specific index $p$ in eq. (\ref{n_E}). We ignored this terms for our order of magnitude estimate because it is  proportional  to $T^{-1}_{\rm out}$ term which 
  we expect to be  cancelled with good accuracy by $T^{-1}_{\rm in}$ as argued above.
  
  This dramatic simplification allows us to estimate the ratio $r$ of the positrons which can get accelerated to sufficiently high energy $E\in (0.5-1)$ MeV  such that these positrons may become the seed particles\footnote{\label{acceleration}In our arguments we assume that the efficiency of acceleration is sufficiently high (the so-called first-order Fermi acceleration) such that $\ln \beta\sim v_{\rm AQN}/c$. In this case the energy $E(t)$  can indeed reach the MeV range on time scale of few ms according to (\ref{E_t}) as $\tau_{e^+}$    is measured in $\mu s$, see (\ref{tau+}). The ms time scale for the acceleration is consistent with this proposal as the AQN propagates only few hundred meters during this ms-interval    and remains deep  inside the region $\sim L_{\cal{E}}$ where electric field $\cal{E}$ should be  large according to basic requirement ``b" as listed at the very beginning of section  \ref{sect:thunder}.} according to criteria (\ref{E_c}).
  With these assumptions the ratio $r$ is estimated as follows:
  \be
  \label{r}
  r\approx \frac{\int_{0.5 MeV}^{1 MeV} E^{-1} dE}{\int_{10~keV}^{MeV} E^{-1} dE}\sim \frac{\ln2}{\ln 10^2}\sim 0.15,
  \ee
  where with logarithmic accuracy we integrated only over energies $E\in (0.5-1)$ MeV because the relativistic positrons with energy $E\gtrsim  \rm MeV$ do not suffer much from re-scattering losses 
  as the cross section $\sigma$ becomes sufficiently small and the mean free path for energetic positrons (\ref{e+}) exceeds the scale $l_a$. 
  Therefore, the dominant portion of the positrons  with  MeV energy will further get accelerated to $10~\rm  MeV$  very quickly   in fraction of $\mu s$ according to (\ref{E_MeV}). 
  
  To conclude: the positrons which are initially injected into the system with typical energy $\sim 10$ keV will get accelerated to $\sim 10$ MeV energy 
  and could play a role of seed particles triggering a very powerful lightning strike as argued in Sect. \ref{fate-positrons}.
  
 \subsection{Shock Acceleration: application to the   electrons}\label{shock-electrons} 
 The acceleration of the AQN-induced electrons by the same mechanism 
 has dramatically different outcome in comparison with the case of positrons considered above.
 The crucial difference is related to the fact that the electrons are produced with typical energy $\sim \rm eV$ as explained in Sect. \ref{fate-electrons},  in contrast with   typical energy of the AQN-induced    positrons which is order of $T\sim 10$ keV. These free thermal electrons  lose its energy within a very short period of time of $\sim 10^{-8} \rm s $ and    disappear   rather rapidly on the time scale $\Delta t_e \sim 10^{-7} \rm s $ (attachment time) as a result of interaction with air molecules,  see  \cite{Gurevich_2001}. Therefore, it is highly unlikely that the AQN-induced eV energy electrons could accelerate to the relativistic energies by the Fermi mechanism, in contrast  to positrons considered above. Therefore, the AQN-induced electrons  cannot serve as seed particles, in contrast with positrons. 
 Nevertheless, these AQN-induced electrons could  play a key dynamical role in the lightning strike  as a very large number of electrons almost instantaneously   get injected into the system,   as equations  (\ref{injected}) and (\ref{e-fold})   suggest. 
 
  \subsection{Diffusion of injected electrons}\label{diffusion}
  
 Another way to represent the powerful jolt reflected by (\ref{injected})    is to estimate the electron number density $n_e(r)$ in close vicinity of the nugget
 which is given by (\ref{e-density}). However, this estimate assumes that the diffusion of the 
 emitted electrons is negligible. In this Appendix we justify this assumption by estimating diffusion features of the emitted electrons. 
 \exclude{
 A proper way to study this problem  is of course to analyze the corresponding kinetic equation accounting for the diffusion, collisions, and the source of the low energy electrons in form  (\ref{collisions}) for the rate of injection of the  low energy AQN-induced electrons   resulting from the   collisions with atmospheric molecules. 
 However, for an order of magnitude estimate one can take a ``snapshot" of this complicated dynamics and estimate average density $\la n_e(r)\ra $ in close vicinity of the nugget  assuming that the injection is instantaneous. 
 }
For the estimate we  use  conventional expression  for the Green's function $G(t, \mathbf{r})$ which assumes the standard form:  
 \be
 \label{n_e-condition}
 G(t, \mathbf{r})=\frac{1}{(4\pi D t)^{3/2}} \exp{\left(-\frac{\mathbf{r}^2}{4Dt}\right)}~\rightarrow~ \la  \mathbf{r}^2\ra\approx 6 Dt~~~~~~~~~~
 \ee
 where $D$ is the diffusion coefficient which has dimensionality $\rm (cm^2/s)$ and will be numerically estimated below. In estimate (\ref{n_e-condition}) it is assumed that
 the diffusion is spherically symmetric process, which is justified if typical velocities of the emitted electrons $v_e$   are much greater than the velocity of the nugget, i.e.  $c\gg v_e\gg v_{\rm AQN}$. 
 \exclude{To simplify estimates further we also assume that the injection occurs instantaneously  at a  well localized point such that the density distribution at $t=0$ is
 \be
 \label{t=0}
 n_e(t=0)= [N^{e^-}] \delta(\mathbf{r}), ~~~ [N^{e^-}] \sim 10^{11},  
 \ee
 where $N^{e^-}\sim 10^{11}$ is the number of electrons injected into the system during the life-time of electrons $\Delta t_e$ as estimated in (\ref{injected}).
With these simplifications 
the distribution at later times $t>0$  is given by
\be
 n_e(t, \mathbf{r})=\frac{[N^{e^-}] }{(4\pi D t)^{3/2}} \exp{\left(-\frac{\mathbf{r}^2}{4Dt}\right)}.
\ee
  It is convenient to express  this  distribution in terms of $\la  \mathbf{r}^2\ra\approx 6 Dt$ as defined by (\ref{n_e-condition}):
  \be
  \label{n_e1}
 n_e(\mathbf{r})=[N^{e^-}]\cdot \left( \frac{ 3}{2\pi \la  \mathbf{r}^2\ra} \right)^{\frac{3}{2}}\cdot \exp{\left(- \frac{3\mathbf{r}^2}{2\la  \mathbf{r}^2\ra}\right)}.~~~
\ee
 }
 
 For the numerical  estimation of  the diffuse coefficient $D\equiv v_e^2/(2\nu)$  we use  $\nu\sim  10^{11} s^{-1}$ where $\nu$ is  the collision frequency, see  e.g. \cite{Gurevich_2001}. The velocity $v_e$ of low energy electrons can be estimated as 
 $v_e/c\simeq \sqrt{2 E_e/m} \sim 0.6\cdot 10^{-2}$ for $E_e\sim 10$ eV.  Collecting all these  factors together we arrive to the following numerical estimates for  the diffuse coefficient $D$ and    the maximal distance from the point of emission $ \la  \mathbf{r}^2\ra_{\rm max} $: 
 \be
 \label{numerics} 
 D\sim 3.6\cdot  10^5 {\rm \frac{cm^2}{s}},  ~~\la  \mathbf{r}^2\ra_{\rm max} \sim 6D (\Delta t_e)\sim 0.04 ~{\rm cm^2}.~~~~~~~
 \ee
This estimate unambiguously shows that the diffusion is relatively minor effect as $\la  \mathbf{r}^2\ra_{\rm max} \ll \bar{r}^2$ where $\bar{r}\sim 3 \rm ~cm$ enters the estimate (\ref{e-density}).

\exclude{
where  $ \la  \mathbf{r}^2\ra_{\rm max}$ is defined as $ \la  \mathbf{r}^2\ra$ at $\Delta t_e \sim 10^{-7} \rm s $ (attachment time)   when the electrons start to disappear from the system. 
At this time new injection occurs with the same intensity (\ref{t=0}),
 and next cycle repeats itself \footnote{The process of electron's injection resulting from ionization as explained in Sect. \ref{fate-electrons} is of course a continues process with the rate (\ref{collisions}). We represent the injection in form of short periodic pulses exclusively  to simplify  our estimates.  Our goal is to produce   average characteristics of the electron density,  which should not be very sensitive to the   spectral characteristics  of the pulses, but only to their intensity. This insensitivity is in fact tested by another estimate (\ref{e-density})  which is not based on this model with injection in form of short pulses, but perfectly consistent with (\ref{n_e}) and (\ref{n_e2}). The motion of the AQN also does not modify our estimate as the AQN traverses   a short distance $\Delta d\simeq v_{\rm AQN}\Delta t_e\sim 3\rm cm$ during $\Delta t_e$.}.

 Using these numerical parameters  one can estimate   an average   electron number density $\la n_e(r)\ra $ as follows:
 \be
 \label{n_e}
 \la n_e\ra \sim [N^{e^-}] \cdot \left[\frac{3}{ 4\pi\la  \mathbf{r}^2\ra^{3/2}_{\rm max}} \right] \sim  \frac{ 10^{9}}{\rm cm^3}. 
 \ee
 Another characteristic is  $n_e(\mathbf{r}=0)$ computed at the same time corresponding to the same $\la  \mathbf{r}^2\ra_{\rm max}$.
 From (\ref{n_e1}) we have
 \be
\label{n_e2}
 n_e(\mathbf{r}=0)=[N^{e^-}]\cdot \left( \frac{ 3}{2\pi\la  \mathbf{r}^2\ra_{\rm max}} \right)^{\frac{3}{2}}  \sim  \frac{ 10^{9}}{\rm cm^3},~~~~~
\ee
  which assumes a similar numerical value as (\ref{n_e}). The both estimates (\ref{n_e}) and (\ref{n_e2}) are consistent with relation   (\ref{e-density})  which is not based    on the  model when the  injection is represented by   short pulses. This consistency  justifies our simplified diffuse model  for estimations of the average characteristics of the system. 
 
 One should also note that the electron's density decays very fast at large distances at $r \gtrsim \sqrt{\la  \mathbf{r}^2\ra_{\rm max}}\approx 3.1~ {\rm cm}$
 from the nugget as equation (\ref{n_e1}) explicitly shows.  This feature is related to very short  life time (attachment time) of free electrons and incorporated into our diffuse  model when injection is parameterized  by a short pulse.  
 }

\section*{References}
   \bibliographystyle{utphys}
   
  \bibliography{AUGER}

\end{document}